\documentclass[11pt]{article}
\pdfoutput=1 

\usepackage{jheppub} 

\usepackage[T1]{fontenc} 

\newcommand{\ket}[1]{|{#1}\rangle}

\newcommand{\bra}[1]{\langle{#1}|}

\newcommand{\ev}[1]{\langle{#1}\rangle}

\newcommand{\R}{\mathbf{R}}
\newcommand{\Z}{\mathbf{Z}}

\newcommand{\del}{\partial}

\DeclareMathOperator{\Tr}{Tr}
\DeclareMathOperator{\sgn}{sgn}

\DeclareMathOperator{\arccot}{arccot}
\DeclareMathOperator{\diag}{diag}
\DeclareMathOperator{\Hspan}{span}

\title{Disk entanglement entropy for a Maxwell field}

\author[a]{C\'esar A. Ag\'on,}
\author[a]{Matthew Headrick,}
\author[b]{Daniel L. Jafferis,}
\author[a]{and Skyler Kasko}

\affiliation[a]{Martin Fisher School of Physics, Brandeis University, Waltham, Massachusetts, USA}
\affiliation[b]{Center for the Fundamental Laws of Nature, Harvard University, Cambridge, Massachusetts, USA}

\abstract{In three dimensions, the pure Maxwell theory with compact $U(1)$ gauge group is dual to a free compact scalar, and flows from the Maxwell theory with non-compact gauge group in the ultraviolet to a non-compact free massless scalar theory in the infrared. We compute the vacuum disk entanglement entropy all along this flow, and show that the renormalized entropy $F(r)$ decreases monotonically with the radius $r$ as predicted by the F-theorem, interpolating between a logarithmic growth for small $r$ (matching the behavior of the $S^3$ free energy) and a constant at large $r$ (equal to the free energy of the conformal scalar). The calculation is carried out by the replica trick, employing the scalar formulation of the theory. The R\'enyi entropies for $n>1$ are given by a sum over winding sectors, leading to a Riemann-Siegel theta function. The extrapolation to $n=1$, to obtain the von Neumann entropy, is done by analytic continuation in the large- and small-$r$ limits and by a numerical extrapolation method at intermediate values. We also compute the leading contribution to the renormalized entanglement entropy of the compact free scalar in higher dimensions. Finally, we point out some interesting features of the reduced density matrix for the compact scalar, and its relation to that for the non-compact theory.}

\preprint{BRX-TH-672}

\begin{document}
\maketitle
\flushbottom

\section{Introduction}

\subsection{Motivation}

It has recently become clear that the renormalized disk entanglement entropy (EE) can serve as a useful tool for understanding renormalization-group flows in three-dimensional theories. Defined by
\begin{equation}
F(r):=rS'(r)-S(r)\,,
\end{equation}
where $S(r)$ is the vacuum EE for a disk of radius $r$ on the plane, this function derives its power from the following three properties, which are believed to hold in any unitary Lorentz-invariant theory:
\begin{itemize}
\item It is ultraviolet-finite and scheme-independent (unlike $S(r)$, which is defined only in the presence of a UV cutoff) \cite{Liu:2012eea}.
\item For a conformal theory, it is constant and equal to the renormalized three-sphere free energy $F_{S^3}=-\ln Z_{S^3}$ \cite{Casini:2011kv}.
\item It is monotonically decreasing (the so-called F-theorem) \cite{Myers:2010xs,Myers:2010tj,Jafferis:2011zi,Casini:2012ei}.
\end{itemize}
These properties make $F(r)$ closely analogous to C- and A-functions in two \cite{Zamolodchikov:1986gt,Cappelli:1990yc} and four \cite{Komargodski:2011vj} dimensions respectively, which are also finite, monotonically decreasing, and equal at a fixed point to a quantity that intrinsically characterizes the CFT, the $a$-type Weyl-anomaly coefficient.

However, there is a crucial difference between $F(r)$ and C- and A-functions. Whereas, at least in a heuristic sense, the latter count the number of local, propagating degrees of freedom at the scale $r$, the same is definitely not true for $F(r)$. Indeed, $F$ is non-zero even in a topological theory. For example, pure Chern-Simons theory has a non-trivial $S^3$ partition function, and moreover there are edge states on the disk; for the abelian theory at level $k$, $F=\frac12\ln k$. This means that $F$ cannot be extracted from local correlation functions, unlike the Weyl-anomaly coefficients in even dimensions. This leads to the question of whether there exist RG flows with more propagating degrees of freedom in the infrared than in the ultraviolet, at the expense of topological ones. For example, the F-theorem would permit a flow from a weakly-coupled, large-level Chern-Simons-matter theory with $F^{\rm UV} \sim \frac{1}{2}\ln k + N_f^{\rm UV}$ to an IR theory with $N_f^{\rm IR} > N_f^{\rm UV}$, as long as  $N_f^{\rm IR} <F^{\rm UV}$. It would be of great interest either to find examples of such exotic RG flows, or to rule them out.

In this paper, we will study another case that dramatically illustrates the failure of $F(r)$ to count local degrees of freedom. The example is the pure Maxwell theory with compact $U(1)$ gauge group. Although this theory contains only a single field, by adding a Chern-Simons term and appealing to the F-theorem we can argue that $F$ becomes arbitrarily large in the ultraviolet. Consider the Chern-Simons-Maxwell theory at level $k$. At energies far above the mass scale $k g^2$, where $g$ is the gauge coupling, the system behaves like the free Maxwell theory, while below that scale it is gapped and only the topological Chern-Simons theory remains. Therefore the F-theorem implies that the UV value of $F(r)$ for Maxwell must be larger than $\frac12\ln k$, for any $k$; in other words, it is divergent. More quantitatively, at weak coupling, hence large $k$, one expects a matching  at the crossover scale $r_{\rm cross} = 1/k g^2$: $F(r_{\rm cross}) \sim \frac{1}{2} \ln k$. This suggests $F(r) \sim -\frac{1}{2} \ln (r g^2)$ for small $r$. This behavior is related to the fact that the Maxwell theory with non-compact gauge group, which is the UV fixed point, is scale- but not conformally invariant \cite{ElShowk:2011gz,Jackiw:2011vz,Dymarsky:2013pqa}, and therefore needn't have a finite three-sphere free energy. If $F(r)$ indeed becomes arbitrarily large in the UV, then we have the question of whether, by a small deformation, this theory containing a single field can be made to flow to one with an arbitrarily large number of propagating fields.

On the other hand, this theory has a more conventional behavior in the IR, which we can most easily understand by dualizing it to a scalar. As we will review in section \ref{dualityreview}, the scalar is periodic with period equal to the gauge coupling $g$. Being free, the theory flows only classically, and the IR fixed point is simply the ordinary, minimally-coupled scalar (see e.g.\ \cite{Dymarsky:2013pqa}). This theory can be made conformal by adding a conformal coupling, which doesn't affect flat-space properties like the disk EE. Hence we expect $F(r)$ to asymptote at large $r$ to the conformal-scalar free energy $F^{\rm nc}=(\ln 2)/8-3\zeta(3)/(16\pi^2)$ \cite{Klebanov:2011gs}.

In this paper we will therefore explicitly calculate $F(r)$ all along the flow for this theory, in order to confirm these predictions for its UV and IR behavior and to check that it interpolates monotonically between them as predicted by the F-theorem.\footnote{The behavior of the renormalized EE for the periodic scalar in three dimensions was also discussed recently in \cite{Metlitski:2011pr,Grover:2012sp}.}

A different motivation for this calculation is to advance the technology available for calculating EEs in quantum field theories in general. For many reasons, including the ones given above, EEs are interesting and useful probes of field theories, yet they are also notoriously difficult to calculate (except in the holographic context). In this respect, the three-dimensional Maxwell/periodic scalar theory is ``just non-trivial enough'' to allow us to carry the calculation through, but along the way we will have to sharpen our tools, in ways that may be useful for calculations in other theories. In particular, there are two notable advances, which will be discussed at length below, involving: (1) the implications of spontaneous symmetry breaking in applying the replica trick, and (2) the application of numerical extrapolation to extract the von Neumann entropy from the R\'enyi entropies.

There are several generalizations of our work that would be interesting to explore. The first would be to the Chern-Simons-Maxwell theory, to confirm the behavior of $F(r)$ predicted above. Another would be to the Yang-Mills theory.\footnote{We thank I. Klebanov and S. Pufu for pointing out this possibility to us.} In the $SU(N)$ theory, for example, presumably $F(r)\sim-\frac12(N^2-1)\ln(rg^2)$ for small $r$, since in the UV it just looks like $N^2-1$ free abelian theories, while for large $r$, $F(r)\to0$. However, calculating $F(r)$ all along the flow might be quite challenging. A generalization in a different direction would be to calculate the ball EE of a periodic scalar in higher dimensions, which is dual to a higher-form gauge field; while we will not do this all along the flow, we will discuss the small-$r$ behavior below.

\subsection{Results}\label{sec:results}

In Section \ref{calculation}, we will calculate $F(r)$ all along the flow. All of the above expectations are confirmed by our calculation: $F(r)$ is monotonically decreasing, and for small and large $r$ we have
\begin{equation}\label{Fsmallr}
F(r) \approx \begin{cases}F^{\rm nc} + \frac12\left(1+\gamma-\ln(rg^2)\right),& rg^2\ll1\\
F^{\rm nc}\,,&rg^2\gg1
\end{cases}
\end{equation}
($\gamma$ is the Euler-Mascheroni constant). A plot of $F(r)$ is shown in figure \ref{fig:Fplot}. We carry out the computation using the replica trick, so along the way we also find an explicit analytic expression for the R\'enyi entropy $S_n(r)$ for integer $n>1$.

The IR and UV behaviors can be understood qualitatively using the description in terms of the periodic scalar, where it is clear that the theory undergoes spontaneous symmetry breaking.\footnote{In the gauge-theory language, the physics looks more complicated, since the spontaneous breaking of the scalar's shift symmetry corresponds to monopole condensation. Note that this is slightly different from the (perhaps more familiar) case in which the symmetry is explicitly broken as the scalar acquires a mass by the Polyakov mechanism.} In the IR, this effect is dominant, the period is effectively large, and so the periodicity is irrelevant. On the other hand, in the UV the period is effectively small, the spontaneous symmetry breaking is irrelevant, and the field fluctuates over the entire circle. This implies that the EE includes a term equal to the logarithm of the volume of the target space, which is just the periodicity $g$. Since $F(r)$ is cutoff-independent, by dimensional analysis $g$ must be multiplied by $r^{1/2}$, i.e.\ $F(r)\sim-\frac12\ln(rg^2)$.

Due to the lack of conformal invariance, the renormalized three-sphere free energy for this theory is required to be neither constant nor equal to the renormalized EE. Nonetheless, it is interesting to compare it to $F(r)$. The free energy was calculated in \cite{Klebanov:2011td} using the gauge-theory description:\footnote{The three-sphere free energy contains both cubically and linearly divergent terms. By ``renormalized free energy,'' we mean the free energy where those divergences are removed by appropriately adjusting the cosmological constant and Einstein terms in the background action. In practice the calculation is performed using zeta-function regularization, which automatically discards these divergences.}
\begin{equation}\label{free}
F_{S^3}(r) = \frac{\zeta(3)}{4\pi^2}-\frac12\ln(rg^2)\,.
\end{equation}
In appendix \ref{freeenergy} we reproduce this result using the scalar description. The second term \eqref{free} can be seen in that description as arising from the integral over the zero mode. Because the scalar is compact, it must be minimally coupled to the sphere metric: no conformal mass can be added. Therefore the constant mode of the scalar is unsuppressed, and the partition function receives a factor of the volume of the field space, which is $g$. As for the EE, by dimensional arguments one must have $r^{1/2}g$, leading to the term $-\frac12\ln(rg^2)$. Similarly, in the Maxwell description one must divide by the volume of the gauge group, which is $1/g$, leading to the same term. While \eqref{free} does not equal $F(r)$, it is interesting that the logarithmic $r$-dependence is the same as the small-$r$ behavior of $F(r)$, \eqref{Fsmallr}. Since there is no canonical way to relate the disk radius to the three-sphere radius, we cannot meaningfully compare the constant terms in the two formulas. Nonetheless, apparently we still have something like $F_{S^3}(r)=F(r)$ in the UV, despite the fixed point not being conformal. This may be viewed simply as a coincidence arising from dimensional analysis. However, it would be interesting to see if there is a generalization of the arguments of \cite{Casini:2011kv} to scale- but not conformally-invariant theories, which could apply more generally.

The physics of the compact free scalar theory in any dimension greater than 2 is similar. The field undergoes spontaneous symmetry breaking, thus at long distances, $S(r)$ approaches that of the ordinary free scalar, while at short distances the compactification is important, and in section \ref{sec:higherd} we derive an additional logarithmic contribution of $-\frac12\ln(r^{d-2}g^2)$.

\subsection{Method}

The calculation of the disk EE was carried out using the replica trick. Consider a
semiclassical wavefunctional basis for the Hilbert space on the
disk, $\Psi(\phi) = \langle \phi | \Psi \rangle$, where $\phi$ are
the field values on the disk. Then the reduced matrix of the
vacuum can be expressed in terms of $\langle \phi | \rho |
\phi' \rangle$, given by the path integral in $\R^3$ where the
field approaches $\phi,\phi'$ as one approaches the disk from above
and below respectively, and with appropriate boundary conditions at infinity
(normalized by the vacuum path integral, $Z_1$).

Therefore, $\Tr \rho^n = Z_n/Z_1^n$ is given by the
partition function on a $n$-sheeted cover of $\R^3$, branched over
the disk. Defining the R\'enyi entropy $S_n = -\ln\Tr \rho^n/(n-1)$, the
extrapolation to $S_1$ is the von Neumann EE. It
would be interesting to understand what the analyticity
properties of $S_n$ are, which should allow a unique continuation to
$n=1$. (See the discussion in \cite{Calabrese:2009ez}.) In this paper, we will assume that such a unique continuation exists.

For the compact scalar theory of interest in this paper, the result takes the form of an instanton sum, leading to an expression for the
$S_n(r)$ in terms of Riemann-Siegal theta functions given as a sum
over $\Z^{n-1}$. It is not known how to analytically continue theta
functions in the argument $n$. In the limits of small and large $r$, we will be able to express $S_n(r)$ in a simpler form that will allow a straightforward analytic continuation. For intermediate values of $r$, we will apply a numerical extrapolation technique. Specifically, for a given $r$ we approximate $S(r)$ as $f_{(p,q)}(1)$, where $f_{(p,q)}(n)$ is the rational function of degree $(p,q)$ defined to fit $S_n(r)$ for $n = 2, \dots , p+q+2$. We find that this technique
gives much better results than polynomial extrapolation. Since it
has not before been applied in a similar situation, we make
various checks of the convergence.

The distinction between the R\'enyi entropies of the compact and
ordinary free scalar theories derives from an intriguing interplay
between short and long distances. Due to the IR symmetry breaking,
the field must go to the same value in all
of the $n$ asymptotic regions. In the circle valued theory, this
results in a sum over instanton winding sectors. We will see that
this gives rise to the logarithmic behavior of the entropies in
the UV.

In each winding sector, one must find the appropriate saddle point for the scalar field. Since the theory is free, the fluctuation determinant will be independent of the instanton sector, and we will find $$Z_n = Z_n^{\rm nc} \sum_{w \in \Z^{n-1}} e^{-wM_n w},$$ for an appropriate matrix $M_n$. 

Determining this matrix involves finding the solution to the Laplace equation on the $n$-sheeted cover of $\R^3$ with asymptotics given by $w$. As for the compact scalar in two dimensions \cite{Calabrese:2009ez}, it is useful to write this as an $\R^n$ valued field on the base $\R^3$ with monodromy through the disk given by the clock matrix ($\phi_i \rightarrow \phi_{i+1}$). 

One can diagonalize the monodromy in a complex basis, leading one to the problem of solving the Laplace equation for a complex scalar on $\R^3$ that goes to 1 at infinity and has a multiplicative phase cut across the disk. Using numerical methods, we find strong evidence for an exact closed form expression for the on-shell action of this solution, thus determining the matrix $M_n$.

\subsection{Effects of discrete gauging}

In the superselection sector of the vacuum in Minkowski spacetime,
there is no difference between the Hilbert spaces of the compact
and ordinary scalar theories. The difference between their
entanglement entropies results from a distinction in how the
Hilbert space decomposes into regions. In particular, on the disk
the compact theory has fewer states, since the wavefunctions must
be periodic under the shift.

An interesting point in this context is that in continuum quantum
field theory the Hilbert space in a region is only defined once
appropriate boundary conditions are imposed. In other words,
different boundary conditions are associated to distinct
superselection sectors. Of course, in the continuum theory, the
field values must be continuous across the boundary of the disk, so
factorization of the Hilbert space is violated; this is why the
entanglement entropy is always defined as a limit of a quantity on
the lattice.

In section \ref{sec:structure}, we show that this implies that in free field theory,
the density matrix is block diagonal in the basis of field values
at the boundary, i.e.\ $[\rho, \phi|_{\del D}] = 0$. More generally, we
show that in any conformal field theory in the conformally-invariant vacuum and for a disk-shaped region, $[\rho, {\cal
O}|_{\del D}] = 0$ for any scalar primary operator ${\cal O}$.

This implies that the density matrix in the compact scalar theory
is formed from a state that is summed over the $\Z$ shifts of the
circle identification as compared to the ordinary scalar theory.
This results in a {\it purer} density matrix, smaller $S_n$ for all $n$, and
thus the larger $F$ that we find.

It might seem strange that a quantity of interest to RG flows in
flat space should depend on more data than the Hamiltonian and
Hilbert space of the theory in Minkowski spacetime. However, the
compact and ordinary scalar theories do differ in the spectrum of
local operators (the collection of {\it all} operators acting
abstractly on the Hilbert space is the same).

In an otherwise identical theory with fewer operators regarded as
local, more couplings to other fields are allowed, which would
have not been local otherwise. Thus it makes sense in principle
that validity of the $F$ theorem requires that $F$ be larger for the compact scalar theory. It would be extremely interesting to find such RG
flows that involve couplings to fields that are not mutually local
with respect to $\phi$, but are with respect to $\del_\mu \phi$
and $e^{2\pi i\phi/g}$.

\subsection{Factorization and the definition of entanglement entropy}

The calculations of EEs in this paper, as in most of the literature on EEs in non-holographic field theories, rely on the replica trick, which reduces them to calculations of the partition functions on certain manifolds, which are rather standard field-theory quantities. Nonetheless, we would like to make a few comments about the definition of the EE in field theories, which is a somewhat more novel type of quantity.

The definition of EE that is usually found in the literature starts with the assumption that the Hilbert space factorizes according to the spatial regions involved, for example for the disk $D$, \ $\mathcal{H}=\mathcal{H}_D\otimes\mathcal{H}_{D^c}$. As discussed in the previous subsection (and in more detail in subsection \ref{sec:compactscalar}), this assumption fails in the scalar theory---and presumably in any continuum field theory---due to the requirement of continuity of the field. Since EEs are normally defined in the presence of a cutoff, such as a lattice, one might not be so concerned about this failure of factorization.

However, in the presence of a gauge symmetry, factorization fails even on the lattice, due to the necessity of dividing the configuration space by gauge transformations that act simultaneously on both regions. In other words, the Gauss law links the states in the two regions. This is true both on the lattice and in the continuum. In order to circumvent this issue, one works in a larger Hilbert space in which the Gauss law is relaxed along the entangling surface (see for example \cite{Donnelly:2011hn} and reference therein). This larger space does factorize, and the physical Hilbert space sits inside it. (A simple example is the Kitaev model \cite{Kitaev:1997wr}, which can be viewed either as a $\Z_2$ lattice gauge theory, in which case the Gauss law is imposed at the level of the Hilbert space, or simply as a spin system, where the constraint is added as a term to the Hamiltonian.) It is also important that the larger Hilbert space has a positive-definite inner product, which can be seen by working in temporal gauge (where the time component of the gauge field is set to zero). Factorization and positivity of the larger Hilbert space are important because these properties are assumed in the proofs of important properties of EEs, such as strong subadditivity.

The Gauss law has an interesting consequence for the structure of the reduced density matrix, which is similar to that discussed above for the scalar theory. Consider the Maxwell theory for definiteness. The Gauss law requires that $E_\perp$, the component of the electric field normal to the entangling surface, be continuous across it. This implies that $E_\perp$ must have a definite value in each eigenstate of the reduced density matrix, i.e.\ $[\rho,E_\perp]=0$. When quantizing the Maxwell theory on the disk, one consistent set of boundary conditions (corresponding to an insulating boundary with a fixed charge distribution) fixes $E_\perp$ along the boundary. In this quantization, different charge distributions are different superselection sectors. Apparently the reduced density matrix corresponds precisely to this quantization.

For the periodic scalar, which can be viewed as a gauging by a $\Z$ shift symmetry of the ordinary scalar, the situation is less clear. We show in subsection \ref{sec:compactscalar} that the replica trick correctly reproduces the R\'enyi entropies defined by working in the Hilbert space and carefully taking into account the gauge symmetry. However, the continuity requirement on the field plays a crucial role in that analysis. It is not clear either how to implement the gauge symmetry in a lattice version of the theory, or how to construct Hilbert spaces that obey factorization.

All of these issues point to the desirability of finding a uniform definition of EE that would be valid for a general abstract quantum field theory---without reference to any particular Lagrangian or lattice model---and that wouldn't require special rules for gauge theories, non-trivial target spaces, and the like. In some sense the replica trick provides such a definition, insofar as it reduces the calculation of EEs to that of partition functions, but it is somewhat indirect; in particular, fundamental properties of EE like strong subadditivity are obscured.

\section{Review of Maxwell-scalar duality}\label{dualityreview}

In three dimensions, the Maxwell field may be dualized to a scalar, defined by $\del_\mu \phi = \epsilon_{\mu \nu \rho} F^{\nu \rho}$. This contains all gauge-invariant information about the Maxwell field. Moreover, in spacetimes with non-trivial topology, the free Maxwell theory with compact $U(1)$ gauge group involves a choice of gauge bundle. This is consistent with the $\phi$ scalar theory if one takes the field to be circle-valued, $\phi \equiv \phi + g$. The shift symmetry of the scalar is associated to the current $j_\mu = \del_\mu \phi = \epsilon_{\mu \nu \rho} F^{\nu \rho}$, the topological current of the Maxwell theory.

Furthermore, even in $\R^{2,1}$, the Maxwell theory admits disorder operators around which $\int F \neq 0$. In the theory with compact gauge group, this flux is quantized. These correspond to $e^{2\pi i k\phi/g}$ in the dual photon description. Note that the Hilbert spaces of the compact and non-compact Maxwell theory in $\R^{2,1}$ are identical---the only difference is whether such monopole operators (which obviously can be defined in both cases, as formal operators on the Hilbert space) are considered to be local operators.

Once such operators are regarded as local operators, the vacuum
must be such that they satisfy clustering. This requires that in
the superselection sector of the vacuum the field $\phi$ takes a
definite asymptotic value, breaking the shift symmetry. In other
words, there is a nonvanishing expectation value for the monopole operator in
the vacuum. In the scalar language, this implies that the compact
scalar flows to the ordinary non-compact scalar in the IR.

The Hilbert space on $\R^{2,1}$ of the compact and non-compact scalar theories are also identical. Again, the difference is whether the operators $\int_\gamma dx^\mu (\del_\mu \phi)$ for a path $\gamma$ from a point $p$ to $\infty$ are considered to be local operators.


Another way of viewing the compact scalar theory is to start with
the ordinary free scalar, and gauge the shift symmetry associated
to the current $j_\mu = \del_\mu \phi$ with a $\Z$-gauge field. In
three dimensions, a $\Z$ gauge field can be described by adding a
BF coupling to an $\R$ gauge field, $\int F_A \wedge B$, where
$B$ is a compact $U(1)$ gauge field \cite{Maldacena:2001ss}. This is analogous to the
description of $\Z_k$ valued gauge fields in terms of a BF
action.

 The role of $B$ is to set to 0 the local modes in the
gauge field $A$, so that only the global $\Z$ shifts are gauged.
The resulting action is
$$\int (\del_\mu \phi + g A_\mu)(\del^\mu \phi + g A^\mu) +
i \epsilon^{\mu \nu \rho} A_\mu \del_\nu B_\rho,$$ where the
identification is $\phi \equiv \phi + g$.

One can fix the shift gauge by setting $\del_\mu \phi = 0$. Then
$A_\mu$ can be integrated out, resulting in the free Maxwell
action for the compact $U(1)$ field $B$. From the point of view of the
above action, the UV entanglement entropy results from the $\Z$
gauge fields.

\section{Disk entanglement entropy}\label{calculation}

In this section, we will compute the entanglement entropy $S(r)$ for a disk of radius $r$ on the plane, in the vacuum, for a free compact scalar with periodicity $g$. As reviewed in section \ref{dualityreview}, this theory is equivalent to a pure Maxwell theory with compact $U(1)$ gauge group and gauge coupling $g$. We will employ the replica trick \cite{Callan:1994py,Holzhey:1994we}, which involves first computing the R\'enyi entropies $S_n(r)$ for integer $n>1$ in terms of the partition function on an $n$-fold branched cover of the Euclidean spacetime using the formula
\begin{equation}\label{Renyidef}
S_n(r) = \frac{n\ln Z_1(r)-\ln Z_n(r)}{n-1}\,,
\end{equation}
and then extrapolating in $n$ to obtain the von Neumann entropy $S(r)=S_1(r)$.

Since the theory is free, the calculation of $Z_n(r)$, and hence $S_n(r)$, naturally decomposes into a ``classical'' part involving a sum over classical solutions and a ``quantum'' part involving fluctuations about these solutions. The quantum part is insensitive to the periodicity $g$, and equal to the R\'enyi entropy $S_n^{\rm nc}(r)$ in the non-compact theory, which has been computed previously \cite{Klebanov:2011uf}.\footnote{Actually, the calculation in \cite{Klebanov:2011uf}, which involved a mapping to a branched cover of $S^3$ or $\R\times H^2$, was for a conformally coupled scalar field, whereas the $g\to\infty$ limit of the compact scalar is minimally coupled. However, since the two theories are indistinguishable in flat space, they should have the same disk R\'enyi entropies. The branched cover involved in the calculation of the R\'enyis by the replica trick does have non-zero curvature, so one might be concerned that the two theories might yield different results. However, this curvature is localized along the boundary of the disk and presumably leads only to a shift in the divergent part of the R\'enyi entropies.} These quantities are UV-divergent, taking the form
\begin{equation}
S_n^{\rm nc}(r) = a_n\frac r\epsilon-\gamma_n\,,
\end{equation}
where $\epsilon$ is a UV cutoff, $a_n$ is a non-universal coefficient, and the finite part $\gamma_n$ is universal. Our calculation will focus on the classical part,
\begin{equation}
\Delta S_n(r) := S_n(r) - S_n^{\rm nc}(r)\,.
\end{equation}
The calculation of $\Delta S_n(r)$ for $n>1$ will be done in subsection \ref{sec:Renyis}, and the extrapolation to $n=1$ will be performed in subsections \ref{limits} and \ref{numerical}, first analytically in the limits of large and small $r$, and then using a numerical extrapolation technique for intermediate values. Based on the result, we will also calculate the renormalized EE $F(r):=rS'(r)-S(r)$, as well as its R\'enyi analogue $F_n(r):=rS_n'(r)-S_n(r)$.

In subsection \ref{sec:higherd}, we will briefly discuss the generalization of our results to the compact scalar in higher dimensions, which is dual to a higher-form gauge field.

While our calculation employs the replica trick, it would be very interesting, both for conceptual and technical reasons, to see if the ``real time methods'' that have been successfully employed in numerical calculations of the disk EE for a massive scalar \cite{Liu:2012eea,Casini:2009sr,Huerta:2011qi,Srednicki:1993im} could be adapted to the periodic scalar, and to compare the results to those we obtain here.

\subsection{R\'enyi entropies}\label{sec:Renyis}

To compute the R\'enyi entropies $S_n(r)$ for integer $n>1$, the replica trick instructs us to compute the partition function $Z_n(r)$ on the $n$-fold branched cover of Euclidean $\R^3$, where the branch cut lies on the disk of radius $r$ on the $\tau=0$ plane, where $\tau$ is the Euclidean time direction, and the sheets are connected cyclically across the branch cut. We call this branched cover $E_n(r)$. The R\'enyi entropy is then given by \eqref{Renyidef}.

The action for our scalar is
\begin{equation}
I[\phi] = \frac12\int d^3x\,\partial_\mu\phi\partial^\mu\phi\,.
\end{equation}
A crucial feature of this scalar is that it undergoes spontaneous symmetry breaking. As we will see, this leads to the existence of winding sectors in the path integral, despite the fact that the branched cover $E_n$ has no non-trivial one-cycles. The winding configurations are sensitive to the periodicity of the scalar, and give rise to a $g$-dependence in the R\'enyi entropies. This is in contrast to the situation in two dimensions, where there is no spontaneous symmetry breaking, and hence there are no winding sectors involved in the calculation of the entanglement entropy for a single interval \cite{Calabrese:2004eu}. (On the other hand, for multiple intervals, the branched cover has non-trivial cycles, and the resulting winding sectors lead to a dependence of the R\'enyis on the scalar's periodicity as it does here \cite{Calabrese:2009ez}.)

The calculation of $Z_n(r)$ will proceed in several steps: (1) identify the topologically non-trivial field configurations; (2) separate the classical and quantum contributions to the partition function; (3) factorize the $r$ and $g$ dependence out of the classical action; (4) diagonalize the monodromy on $E_n(r)$ by going to a basis of complex scalar fields; (5) solve the resulting classical electrostatics problem.

\subsubsection{Instanton sum}

Because of the spontaneous symmetry breaking, the scalar field $\phi$ takes on a definite value at infinity. Without loss of generality, let us take this value to be $\phi = 0$. The asymptotic boundary of the branched cover $E_n(r)$ contains $n$ connected components, and on each of them $\phi$ must equal 0.\footnote{We remind the reader that the partition function $Z_n(r)$ on $E_n(r)$ computes $\Tr\rho^n$, where $\rho = \Tr_{D^c}\ket{0}\bra{0}$ and $D^c$ is the complement of the disk $D$. The boundary condition on $\phi$ guarantees that every factor of $\rho$ appearing in $\Tr\rho^n$ arises from the same vacuum $\ket{0}$.} (Topologically, $E_n(r)$ is an $n$-punctured $S^3$.) However, since the field is periodically identified, if we consider a path connecting one asymptotic infinity to another, the field may wind around the circle an integer number of times. By choosing one of the infinities as a base point, it is clear that there are $n-1$ independent winding numbers. Since the field is free, for any given set of winding numbers $w=(w^1,\ldots,w^{n-1})$, there is exactly one solution $\phi_w$ to the classical equations of motion with that particular set of winding numbers. Therefore any field configuration can be uniquely decomposed into a classical part and a quantum fluctuation,
\begin{equation}
\phi = \phi_w+\phi_\text{nc}
\end{equation}
for some $w$, where both components go to 0 at every infinity and $\phi_\text{nc}$ has vanishing winding numbers (and therefore can be thought of as a configuration of the non-compact theory). Correspondingly, the action decomposes,
\begin{equation}
I[\phi] = I[\phi_w] + I[\phi_\text{nc}]\,,
\end{equation}
and the path integral factorizes,
\begin{equation}
Z_n(r) = \int[d\phi]e^{-I[\phi]} = Z^\text{cl}_n(r)Z^\text{nc}_n(r)\,,
\end{equation}
\begin{equation}
Z^\text{cl}_n(r) = \sum_{w\in \Z^{n-1}}e^{-I[\phi_w]}\,,\qquad
Z^\text{nc}_n(r) = \int[d\phi_\text{nc}]e^{-I[\phi_\text{nc}]}\,.
\end{equation}
The non-compact theory does not admit non-trivial classical solutions, so its partition function is simply $Z_n^{\rm nc}(r)$. We thus have
\begin{equation}\label{sclsqu}
\Delta S_n(r) = -\frac1{n-1}\ln Z^\text{cl}_n(r)\,,
\end{equation}
where we have used the fact that there are no winding sectors for $n=1$ so $Z_1^\text{cl}(r)=1$.

In order to compute $Z^\text{cl}_n(r)$, we need to evaluate $I[\phi_w]$. We start by rescaling both the spacetime and the target space in order to scale out the dependence on $r$ and $g$. With $\phi_w(x) = g\hat\phi_w(x/r)$, where $\hat\phi$ is scalar with periodicity 1 living on $E_n(1)$, we have
\begin{equation}
I[\phi_w] = rg^2I[\hat\phi_w]\,,
\end{equation}
where $\hat\phi$ is independent of $g$ and $r$. Since $r$ and $g$ will appear only in the combination $rg^2$, to simplify the notation we will set $g=1$ for the rest of the paper. The factors of $g$ can be restored simply by replacing $r$ by $rg^2$ everywhere.

We now note that $I[\hat\phi_w]$ is a positive-definite quadratic form in the set of winding numbers $w$; this follows from the fact that the action is quadratic in the field, which in turn is linear in $w$: $\hat\phi_{w+w'} = \hat\phi_w+\hat\phi_{w'}$. Hence we can write
\begin{equation}
I[\hat\phi_w] = (M_n)_{jj'}w^jw^{j'}
\end{equation}
for some matrix $M_n$, from which we obtain
\begin{equation}\label{Theta}
Z_n^\text{cl}(r) = \Theta(irM_n/\pi)\,,\qquad
\Delta S_n(r) = -\frac1{n-1}\ln\Theta(irM_n/\pi)\,,
\end{equation}
where $\Theta$ is the Riemann-Siegel theta function
\begin{equation}
\Theta(A)=\Theta(0|A) = \sum_{w\in\Z^{n-1}}e^{\pi iA_{jj'}w^jw^{j'}}\,.
\end{equation}

\subsubsection{Calculation of $M_n$}

Our task is thus to calculate the matrix $M_n$. This involves finding the action for a solution with a given set of winding numbers $w^i$. It is equivalent, and calculationally more convenient, to consider $\hat\phi$ as a non-periodic scalar that asymptotes to 0 on the $n$th sheet and to $w^j$ on the $j$th sheet ($j=1,\ldots,n-1$). Since the equation of motion is just Laplace's equation, we have the textbook electrostatics problem of finding the electrostatic potential with prescribed values on the boundaries of a certain geometry. Furthermore, the action for $\hat\phi$ is just the energy stored in the corresponding electric field. In sum, we are to calculate the capacitance of a certain capacitor---albeit one living in a multi-sheeted geometry.

To avoid having to deal with multiple sheets, it is convenient to diagonalize the monodromy by employing complex linear combinations of the field values on the different sheets (as was done in the two-dimensional case in \cite{Casini:2005rm,Calabrese:2009ez}). Thus, with $y$ denoting a point on $\R^3$ and $x(y,j)$ the corresponding point on the $j$th sheet of $E_n(1)$, we define
\begin{equation}
\tilde\phi_w^k(y) = \frac1{\sqrt n}\sum_{j=1}^ne^{2\pi ijk/n}\hat\phi_w(x(y,j))\,,\qquad (k=1,\ldots,n)\,.
\end{equation}
When crossing the unit disk, the field $\hat\phi_w$ undergoes a monodromy $\hat\phi_w(x(y,j))\to\hat\phi_w(x(y,j+1))$ (where $x(y,n+1):=x(y,1)$), so $\tilde\phi_w^k$ undergoes the monodromy
\begin{equation}
\tilde\phi_w^k\to e^{-2\pi ik/n}\tilde\phi_w^k\,.
\end{equation}
The field $\tilde\phi_w^k$ takes the asymptotic value
\begin{equation}
\tilde w^k = \frac1{\sqrt n}\sum_{j=1}^{n-1}e^{2\pi ijk/n}w^j
\end{equation}
(recall that the asymptotic value on the $n$th sheet has been set to 0, so $j=n$ is not included in the sum). In terms of the fields $\tilde\phi_w^k$, the action becomes
\begin{align}
I[\hat\phi_w] &= \frac12\int_{E_n(1)}d^3x\,\partial_\mu\hat\phi_w\partial^\mu \hat\phi_w \nonumber \\ &=\frac12\sum_{j=1}^n\int_{\R^3} d^3y\,\partial_\mu\hat\phi(x(y,j))\partial^\mu\hat\phi(x(y,j)) \nonumber  \\ &=
\frac12\sum_{k=1}^n\int_{\R^3} d^3y\,\partial_\mu(\tilde\phi^k_w)^*\partial^\mu\tilde\phi^k_w\,.\label{Itemp}
\end{align}

Two observations can be used to simplify \eqref{Itemp}. First, the solution for $k=n$, $\tilde\phi_w^n$, has trivial monodromy and is therefore constant and does not contribute to the sum. Second, we can separate the dependence on the monodromy phase $e^{-2\pi ik/n}$ from the dependence on the asymptotic value of the field $\tilde w^k$ by rescaling $\tilde\phi^k_w$ by the latter. We thus define
\begin{equation}
J(\beta) := \frac12\int d^3y\,\partial_\mu\tilde\phi_\beta^*\partial^\mu\tilde\phi_\beta\,,
\end{equation}
where $\tilde\phi_\beta$ is the solution to Laplace's equation that asymptotes to 1 at infinity and has monodromy across the unit disk
\begin{equation}
\tilde\phi_\beta \to e^{-2\pi i\beta}\tilde\phi_\beta\,.
\end{equation}
We then have
\begin{equation}
I[\hat\phi_w] = \sum_{k=1}^{n-1}|\tilde w^k|^2J(\frac kn)
= \frac1n\sum_{j,j',k=1}^{n-1}w^jw^{j'}e^{2\pi i(j-j')k/n}J(\frac kn)
\,,
\end{equation}
from which we learn that the matrix $M_n$ has components
\begin{equation}
(M_n)_{jj'} = \frac1n\sum_{k=1}^{n-1}e^{2\pi i(j-j')k/n}J(\frac kn)\,,
\end{equation}
i.e.
\begin{equation}\label{Mn}
M_n = U_nJ_nU_n^\dag\,,\qquad (U_n)_{jk}=\frac1{\sqrt n}e^{2\pi ijk/n}\,,\qquad (J_n)_{kk'}=J(\frac kn)\delta_{kk'}\,.
\end{equation}

The problem is thus finally reduced to calculating the function $J(\beta)$, which involves solving an ``electrostatics'' problem for a complex-valued potential on $\R^3$ with a phase monodromy around the unit circle. To avoid distracting from the main line of this paper, we deal with this problem in appendix \ref{Jcalc}, where we find the following formula:
\begin{equation}\label{Jbeta}
J(\beta) = 2\pi(1-2\beta)\tan(\pi\beta)\,.
\end{equation}

Equations \eqref{Theta}, \eqref{Mn}, \eqref{Jbeta} together give an explicit formula for $\Delta S_n(r)$ for arbitrary $r$ and integer $n>1$. The entropies for $n=2,3,4$ are plotted in fig.\ \ref{fig:Splot}. It is worth noting that the formula we've obtained is quite similar in structure to the one obtained in \cite{Calabrese:2009ez} for the R\'enyi entropies of two intervals for a compact scalar in two dimensions, which also involved Riemann-Siegel theta functions taking as their argument an $(n-1)\times(n-1)$ matrix of the form \eqref{Mn}, but with a different function playing the role of $J$. The reason is that the two calculations follow the same outline: the path integral involves a sum over winding sectors, with the on-shell classical action being a quadratic form in the winding numbers, which can be diagonalized by diagonalizing the monodromies.

\subsection{Von Neumann entropy}

Unfortunately, it is not known how to analytically continue a Riemann-Siegel theta function in the dimension of its argument (in this case, $n-1$), so we cannot give an explicit formula for the von Neumann entropy difference $\Delta S(r)$. (The same problem occurs for the two-dimensional compact scalar.) However, in the limits of large and small $r$, we can obtain formulas that can be extrapolated to $n=1$, as we will discuss in subsection \ref{limits}. In subsection \ref{numerical}, we will describe an effective numerical extrapolation method to obtain $\Delta S(r)$ at intermediate values of $r$. To our knowledge, this is the first application of numerical extrapolation to the problem of computing the von Neumann entropy from R\'enyi entropies. It would be very interesting to see if our method can be applied to other cases where it is not known how to perform the extrapolation analytically, such as the two-dimensional compact scalar \cite{Calabrese:2009ez}.

\subsubsection{Infrared and ultraviolet limits}\label{limits}

When $r$ is large, we are exploring the infrared limit of the theory, where it goes over effectively to a non-periodic scalar. Correspondingly, the field configurations with non-zero winding are exponentially suppressed in the path integral, so
\begin{equation}\label{larger}
Z_n(r) \approx Z_n^\text{nc}(r)\,,\qquad S_n(r)\approx S_n^\text{nc}(r)\,.
\end{equation}
Hence $S(r)\approx S^\text{nc}(r)$, so $F(r)$ becomes constant and equal to its value for the non-compact scalar, which was computed using the $S^3$ partition function in \cite{Klebanov:2011gs}:
\begin{equation}\label{largerF}
F(r) \approx F^{\rm nc} = \frac{\ln2}8-\frac{3\zeta(3)}{16\pi^2}\,.
\end{equation}
The corrections to \eqref{larger} are exponentially small in $r$. However, it is difficult to know if that statement survives the extrapolation to $n=1$, and hence also holds for \eqref{largerF}.

On the other hand, in the opposite limit of small $r$, we are exploring the ultraviolet limit of the theory, where it goes over to a Maxwell theory with non-compact gauge group. As the gauge theory becomes classical, the dual scalar becomes strongly coupled, with large topological fluctuations. Since the winding sectors are not very suppressed, we can take the winding numbers $w^j$ to be continuous, and the theta function is well approximated by a Gaussian integral:
\begin{align}
Z^\text{cl}_n(r) &= \sum_w\exp\left(-r(M_n)_{jj'}w^jw^{j'}\right) \nonumber \\
&\approx \int d^{n-1}w\exp\left(-r(M_n)_{jj'}w^jw^{j'}\right) \nonumber  \\ &= \left(\det(rM_n/\pi)\right)^{-1/2} \nonumber  \\ &= (r/\pi)^{(1-n)/2}(\det M_n)^{-1/2}\,.\label{gaussian}
\end{align}
So we have
\begin{equation}
\Delta S_n(r) \approx \frac12\ln(r/\pi)+\frac1{2(n-1)}\ln\det M_n\,.
\end{equation}
From \eqref{Mn} we have
\begin{equation}
\det M_n = \det(UU^\dag)\prod_{k=1}^{n-1}J(\frac kn) = \frac1\pi\left(\frac{4\pi}n\right)^{n-1}\Gamma(\frac n2)^2\,.
\end{equation}
(A short calculation shows that $\det(UU^\dag)=1/n$. The product of the $J(\frac kn)$ can be obtained for odd $n$ with the help of the identity $\prod_{k=1}^{n-1}\tan(\pi k/n)=n\csc(\pi n/2)$, and it can be checked that the resulting formula is correct for $n$ even as well.) Hence
\begin{equation}\label{Snapprox}
\Delta S_n(r) \approx \frac12\ln\left(\frac{4r}n\right)+\frac1{n-1}\ln\left(\frac{\Gamma(\frac n2)}{\sqrt\pi}\right)\,.
\end{equation}
Given this formula, we can now let $n$ be continuous, and take the limit $n\to1$. We find
\begin{equation}\label{Sapprox}
\Delta S(r) \approx \frac12\ln r-\frac\gamma2\,,
\end{equation}
where $\gamma$ is Euler's constant. From this we can also derive $F(r)$:
\begin{equation}\label{Fapprox}
F(r) \approx F^\text{nc} + \frac12+\frac\gamma2-\frac12\ln r = \frac{\ln2}8-\frac{3\zeta(3)}{16\pi^2}+ \frac12+\frac\gamma2-\frac12\ln r\,.
\end{equation}
The sum of the first four terms on the right-hand side has the approximate numerical value 0.852.

A systematic approximation to $Z^{\rm cl}_n(r)$, of which \eqref{gaussian} is the leading term, is obtained by performing a modular transformation on the theta function,
\begin{equation}
Z_n^\text{cl}(r) = \Theta(irM_n/\pi) = \left(\det(rM_n/\pi)\right)^{-1/2}\Theta\left(i\pi M_n^{-1}/r\right),
\end{equation}
and then expanding the new theta function as a sum of exponentials. The corrections to \eqref{Snapprox} obtained in this way are exponentially small in $1/r$. However, just as in the limit of large $r$, it is difficult to extrapolate those corrections to $n=1$, and therefore to know whether the corrections to \eqref{Sapprox} and \eqref{Fapprox} are also exponentially small.

\subsubsection{Numerical extrapolation}\label{numerical}

In principle, the R\'enyi entropies $S_n(r)$ computed for integer $n>1$ in subsection \ref{sec:Renyis} determine them for general real $n\ge0$, and in particular they determine the von Neumann entropy $S(r) = S_1(r)$. However, as mentioned above, it is not known how to describe the analytic function of $n$ that interpolates between the values of theta functions whose arguments are given $(n-1)\times(n-1)$ matrices. The same stumbling block has prevented the calculation of the von Neumann entropy for multiple intervals for a two-dimensional compact scalar, despite the explicit knowledge of the $n>1$ R\'enyis \cite{Calabrese:2009ez}.

In the absence of an analytic technique to perform the extrapolation of the R\'enyis to $n=1$, in this subsection we will test and apply a numerical method. As far as we know, numerical extrapolation has not previously been applied to the computation of entanglement entropies, so we should be careful in choosing our method and checking its reliability. The problem of interpolating or extrapolating a function from its known values at a finite set of points to a new point is of course an old one in numerical analysis, and various techniques have been developed (see chapter 3 of \cite{NR} for an overview). We experimented with several of them, including polynomial interpolation and two types of rational interpolation. We tested each method on two extrapolation problems similar to the one at hand for which the correct values were known a priori:
\begin{enumerate}
\item[(A)] Finding $\Delta S(r)$ in the limit of small $r$ (eq. \eqref{Sapprox}), given $\Delta S_n(r)$ for $n>1$ (eq.\ \eqref{Snapprox}).
\item[(B)] Finding $\Delta S_2(r)$, given $\Delta S_n(r)$ for $n>2$.
\end{enumerate}
We found that the best results were consistently obtained using a simple rational interpolation method, which consisted of fixing the $p+q+1$ independent coefficients of a degree $(p,q)$ rational function so that it passes through $p+q+1$ known data points, and evaluating the resulting function at the new point.\footnote{This method is implemented as the function {\tt RationalInterpolation} in the \emph{Mathematica} package {\tt FunctionApproximations}.} Like Pad\'e approximants, such rational interpolating functions are capable of providing surprisingly accurate approximations---far better than polynomial interpolating functions---especially, as in our case, when used for extrapolation (i.e.\ when the new point is outside the range of values of the known points). The main pitfall with this method is that occasionally---and somewhat unpredictably---the rational interpolating function will happen to have a pole in the vicinity of the new point, yielding a result with a large error. Fortunately, such poles are easy to detect, and can usually be removed simply by slightly changing the degree of the numerator and/or denominator.

\begin{figure}[tbp]
\centering
\includegraphics[width=.6\textwidth]{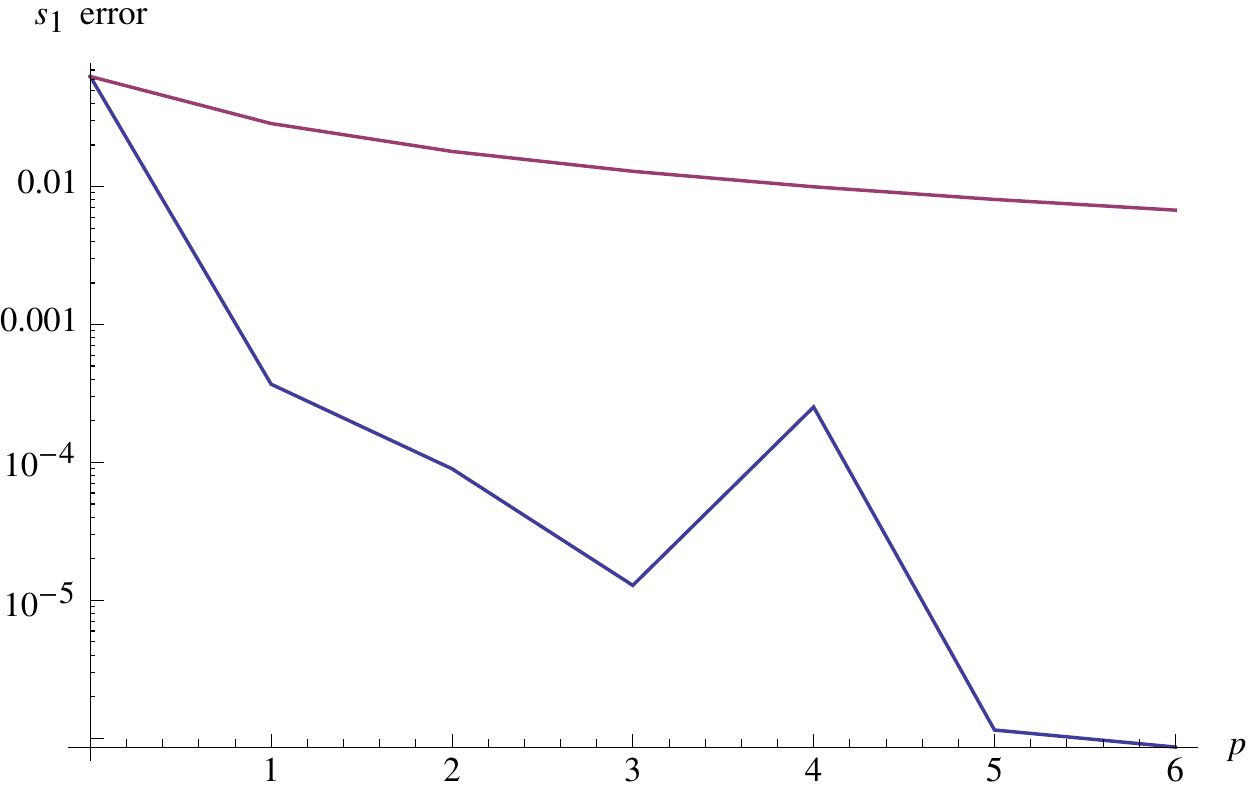}
\caption{\label{fig:s1error}
Absolute value of the error in the predicted value of $s_1$ from (top curve) degree $2p$ polynomial interpolating functions and (bottom curve) degree $(p,p)$ rational interpolating functions, taking as input the values of $s_n$ for $n=2,\ldots,2p+2$.
}\end{figure}

Consider, for example, problem (A) above. We can write \eqref{Snapprox} as
\begin{equation}
\Delta S_n(r) \approx \frac12\ln r+s_n\,,\qquad
s_n = \frac12\ln\frac4n+\frac1{n-1}\ln\left(\frac{\Gamma(\frac n2)}{\sqrt\pi}\right).
\end{equation}
Taking as input the values of $s_n$ for integer $n>1$, we wish to predict the value of $s_1$. From \eqref{Sapprox}, the correct answer is $s_1 = -\gamma/2\approx-0.289$. Figure \ref{fig:s1error} shows the errors in the value of $s_1$ obtained using the degree $2p$ polynomial interpolating function and the degree $(p,p)$ rational interpolating functions, each fit to the values of $s_n$ for $n=2,\ldots,2p+2$. We see that the error generally decreases with $p$ for both types of extrapolation, but far faster for the rational than for the polynomial functions. However, the rational extrapolation seems to have an anomalously large error at $p=4$, relative to the general trend (though still far smaller than for the polynomial interpolating function at the same value of $p$). Inspection of the $(4,4)$ rational interpolating function shows that it happens to have a pole at $n\approx1.04$, leading to a relatively large error in the predicted value at $n=1$. Changing the degree slightly, to $(p,q)=(4,3)$, (3,4), (5,3), or (3,5), removes the pole and reduces the error by a factor of 20. For $p,q$ larger than around 5, the improvements cease to be significant, and in any case it would be impractical for our actual application since the evaluation of the theta functions becomes computationally very expensive for $n$ larger than around 12.\footnote{Incidentally, the $(p,p)$ rational interpolating functions can also be used to predict the value of $s_\infty$, even more accurately than for $s_1$. The error decreases monotonically in $p$, from $7\times10^{-2}$ for $p=0$ to an amazing $4\times10^{-12}$ for $p=5$.}

A similar analysis for problem (B) shows that the rational interpolating functions provide a highly accurate method for predicting $\Delta S_2(r)$ from $\Delta S_n(r)$ for $n>2$ throughout the range of interest of $r$ values (roughly $-0.5<\ln r<2$; for larger and smaller values of $r$ the functions are hardly distinguishable from their corresponding asymptotic approximations). While the most accurate approximation was often obtained using the degree $(5,5)$ rational function, it was found that the degree $(5,4)$ function tended to be less susceptible to the appearance of dangerous poles, and therefore gave more reliable and stable results. (The reasons for this are not clear to us.) The error in $\Delta S_2(r)$ obtained this way ranged between $10^{-9}$ and $10^{-4}$.

\begin{figure}[tbp]
\centering
\includegraphics[width=.65\textwidth]{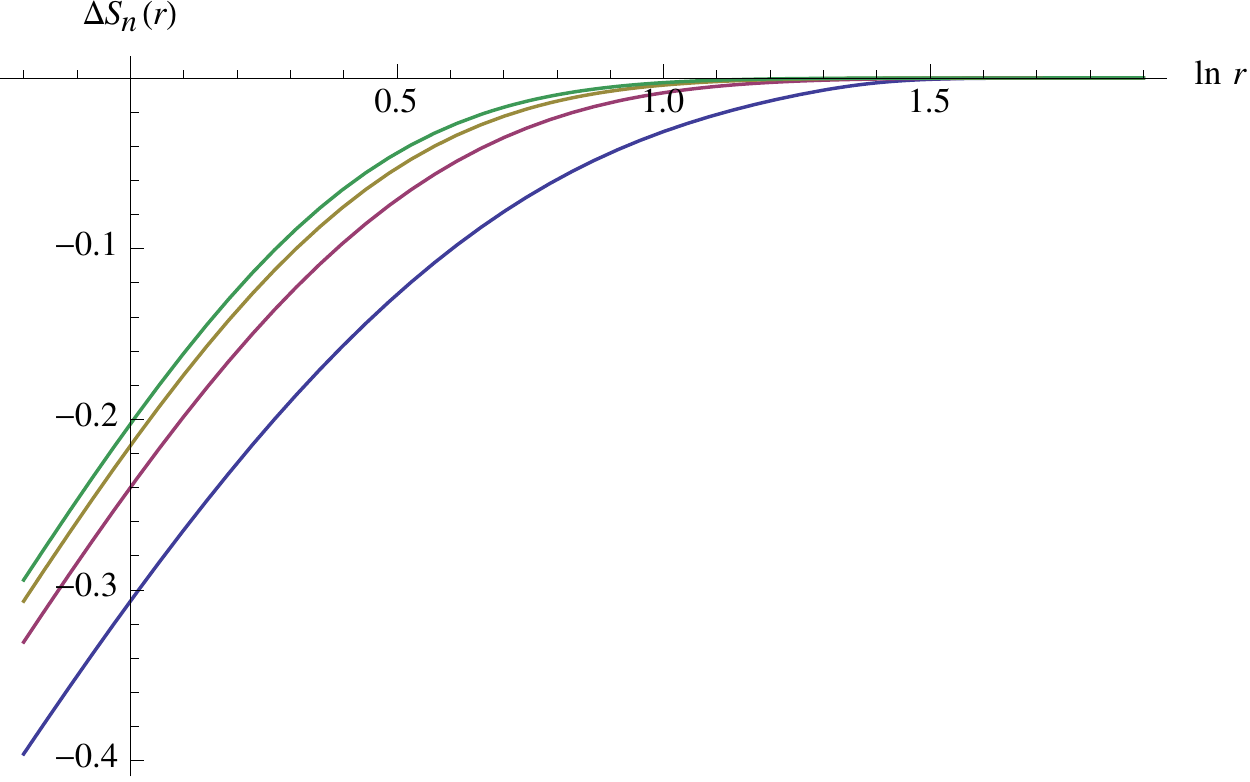}
\caption{\label{fig:Splot} $\Delta S_n(r)$ for $n=1,2,3,4$ (bottom to top). The functions for $n=2,3,4$ are given by equation \eqref{Theta}, while $\Delta S(r)=\Delta S_1(r)$ is obtained by the rational extrapolation method described in the text.}
\end{figure}

Having tested the rational interpolation method, we now turn to its application to predicting the value of $\Delta S(r)$. As with problem (B), we found that the $(5,4)$ rational interpolating function provided a generally very stable approximation, and was used for most data points; however, for values of $r$ where a pole interceded close to $n=1$, the $(5,5)$ approximation was substituted. While rigorous error estimates are difficult to obtain with this method, we believe that the errors should be similar to those found in problem (B) above. The resulting function, along with $\Delta S_n(r)$ for $n=2,3,4$, is plotted in fig.\ \ref{fig:Splot}.

\begin{figure}[tbp]
\centering
\includegraphics[width=.65\textwidth]{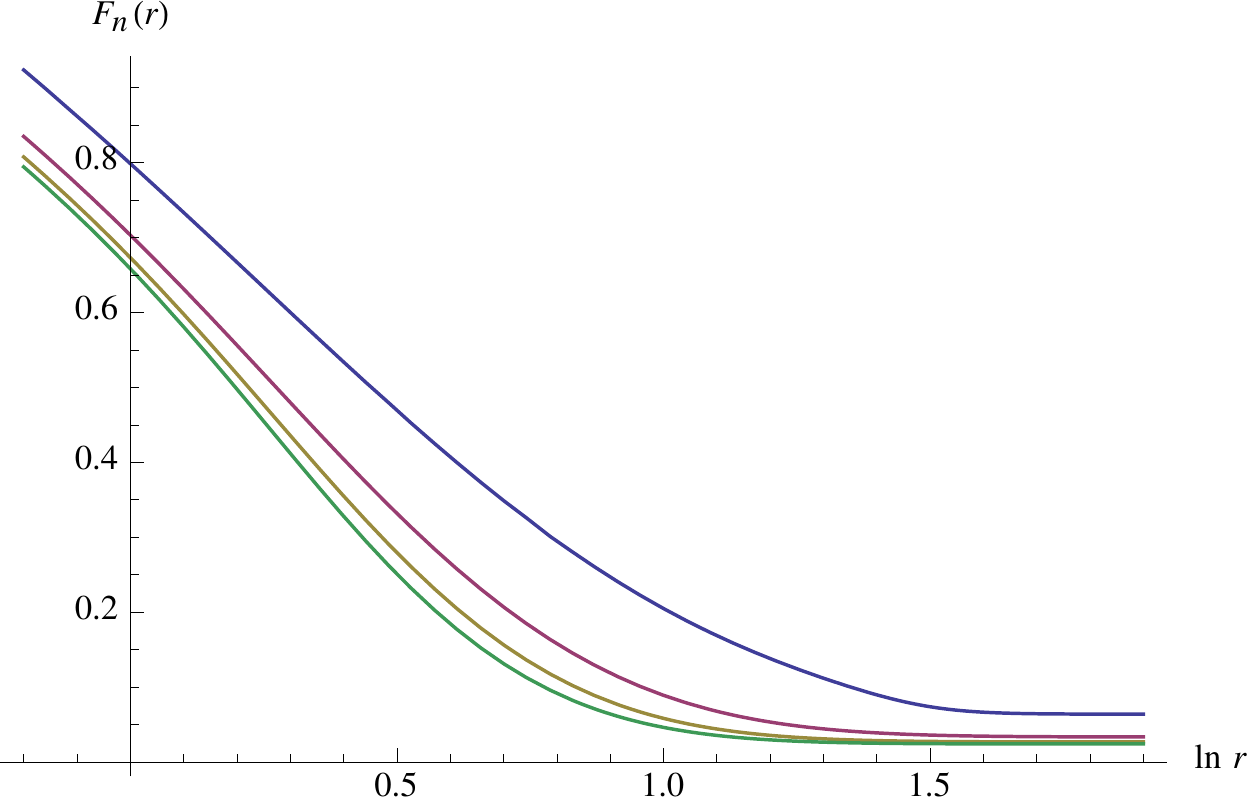}
\caption{\label{fig:Fplot} $F_n(r)$ for $n=1,2,3,4$ (top to bottom). The functions for $n=2,3,4$ are calculated from $\Delta S_n(r)$ using \eqref{Fncalc}, while $F(r)=F_1(r)$ is obtained by the rational extrapolation method described in the text.}
\end{figure}

The renormalized entanglement R\'enyi entropy is defined similarly to the renormalized EE,
\begin{equation}
F_n(r):=rS_n'(r)-S_n(r)\,.
\end{equation}
Like $F(r)$, these functions are UV-finite and constant for CFTs. However, they are not believed to be necessarily monotonic. The values $F^{\rm nc}_n$ for the conformal scalar were computed in \cite{Klebanov:2011uf}. The functions
\begin{equation}\label{Fncalc}
F_n(r) = F^{\rm nc}_n+r\Delta S'_n(r)-\Delta S_n(r)
\end{equation}
are plotted in fig.\ \ref{fig:Fplot} for $n=2,3,4$, along with $F(r)=F_1(r)$, obtained by extrapolation using a (5,4) rational function.\footnote{More precisely, the function $r\Delta S_n'(r)-\Delta S_n(r)$ was extrapolated, and then the constant $F^{\rm nc}$ was added. Incidentally, as another test of the extrapolation method, the value of $F^{\rm nc}$ can be predicted from the values of $F_n^{\rm nc}$ for $n=2,\ldots,11$ using a (5,4) rational function, with an absolute error less than $4\times10^{-5}$.} As predicted by the F-theorem \cite{Casini:2012ei}, $F(r)$ is monotonically decreasing. It is also interesting to note that, for fixed $r$, $F_n(r)$ decreases monotonically with $n$.

\subsection{Compact scalar in higher dimensions} \label{sec:higherd}

The behavior of the ball entanglement entropy for the compact scalar theory in any dimension greater than 2 has a similar structure. This theory is dual to a $(d-2)$-form gauge theory with $U(1)$ gauge group. Again, the vacuum spontaneously breaks the shift symmetry, and one must sum over winding sectors when applying the replica trick. To determine the full $S(r)$ would require finding the appropriate solutions to the Laplace equation, so here we will only calculate the leading logarithmic term at short distances. At long distances, $S(r) \rightarrow S^\textrm{nc}(r)$, reproducing the ordinary free scalar result due to the spontaneous symmetry breaking. 

As in the three-dimensional case, the partition function $Z_n$ must be computed on the $n$-sheeted cover $E_n^d(r)$ of $\R^d$, branched over the sphere of radius $r$. Asymptotically in each sheet, $\phi_w \rightarrow g w_i$, where $g$ is the periodicity of the field and $w_i$ are the winding numbers. In each sector one can find the classical solution $\phi_w$.

The instanton action will be given by $I[\phi_w] = r^{d-2} g^2 I[\hat\phi_w] = (r^{d-2} g^2)\, w M_n^{(d)} w$, where $\hat\phi_w$ is the solution for $g=1$ and $r=1$, and $M_n^{(d)}$ is the $(n-1)$-dimensional quadratic form one would obtain by finding the instanton solutions in the $d$-dimensional case. 

Therefore, in the ultraviolet limit, we can approximate the theta function by a Gaussian integral, which leads to $$Z_n^\textrm{cl}(r) \approx \left(\det \left( \frac{r^{d-2} g^2 M_n^{(d)}}{\pi} \right) \right)^{-1/2}\,.$$ Thus the leading dependence at small $r$ is given by $\Delta S_n (r) \approx \frac{1}{2} \ln(r^{d-2} g^2)$. This implies that 
\begin{equation}\label{UV}
S(r) \approx  - \frac{d-2}{2} \ln(r g^{2/(d-2)})+ S^\textrm{nc}(r)\,. 
\end{equation}

In even dimensions, the entanglement entropy for conformal field theories contains a term that depends logarithmically on the radius, with a coefficient $F$ proportional to the $a$-type Weyl anomaly \cite{Solodukhin:2008dh,Myers:2010tj}. According to \eqref{UV}, the compact scalar in the UV has an additional contribution of $(d-2)/2$, compared to the ordinary free scalar. According to the definition of the renormalized EE $F(r)$ for general dimensions proposed in \cite{Liu:2012eea}, we therefore have
\begin{equation}
F(r)\approx F^{\rm nc}+\frac{d-2}2\,.
\end{equation}
Since the theory flows in the IR to the non-compact scalar, this behavior is consistent with the statement $F^{\rm UV}>F^{\rm IR}$ (even though the theory is not conformal in ultraviolet).

In odd dimensions, $F(r)$ goes to a constant for CFTs, whereas we find that the compact scalar has, in contrast, a divergent value of $F(r)\approx-\frac12(d-2)\ln r$ at short distances.

\section{Structure of the reduced density matrix}\label{sec:structure}

We saw in subsection \ref{sec:Renyis} that $\Delta S_n(r)<0$ for all $n>1$ and all $r$. This was a consequence of the general form of the calculation, and did not depend on the detailed form of the classical solutions involved (see equation \eqref{Theta}). Furthermore, both the analytic and numerical extrapolations in subsections \ref{limits} and \ref{numerical} led to negative values of $\Delta S(r)$. In this section we will put aside the replica trick and try to understand from a more basic point of view why one theory has smaller entanglement (R\'enyi) entropies than the other. The short version of the answer we will find is simply that the compact theory has fewer states, because of the identification $\phi\sim\phi+1$. (As in the previous section, we continue to set $g=1$.) Below we will make this more precise. Many of the remarks below apply more generally than to the problem at hand---vacuum disk entanglement entropy of the compact scalar in three dimensions---but for concreteness we will for the most part restrict our discussion to that case.

The basic story will be as follows. The compact theory can be regarded as a gauging of the non-compact theory under constant shifts of the field. We will show that in the non-compact theory, the reduced density matrix is block-diagonal. The blocks get mapped to each by the gauge transformations, so in the compact theory they are summed. Finally, it is easy to show that, given a set of positive matrices with total trace 1, the sum is purer than the direct sum, and hence has smaller R\'enyi entropies for all $n$ (including $n=1$).

This mechanism is similar to the one that leads to a positive value of $F$ in discrete lattice gauge theories that flow to topological theories in the infrared, such as the Kitaev model \cite{Kitaev:1997wr}. In that case, the Gauss law constraint reduces the number of states appearing in the reduced density matrix of the disk, leading to a constant deficit in $S(r)$ relative to the area-law \cite{2005PhRvA..72a2324H}.

\subsection{Non-compact scalar}\label{sec:noncompactscalar}

We begin with the non-compact theory. Working in the ``position'' basis for the field $\phi$, we label the value of $\phi$ inside the disk $D$ $\phi_D$ and outside it $\phi_{D^c}$. We will denote the reduced density matrix $\sigma$. In this subsection we will show that $\sigma$ is block-diagonal in the value of the field on the edge of the disk, i.e.\ $\sigma$ commutes with the field operator $\hat\phi(x)$ where $x\in\partial D$. This statement will play an important role in understanding the reduced density matrix for the compact scalar in the rest of this section. We will prove the statement in two ways, first using the vacuum wave functional and then by expressing $\sigma$ in terms of conformal generators. We will then give a couple of heuristic ways to understand the statement.

The matrix elements of $\sigma$ are given by tracing the vacuum wave functional $\ev{\phi_D,\phi_{D^c}|0}$ over $\phi_{D^c}$:
\begin{equation}\label{sigmadef}
\ev{\phi_D|\sigma|\phi_D'} = \int[d\phi_{D^c}]\,\ev{\phi_D,\phi_{D^c}|0}\ev{0|\phi_D',\phi_{D^c}}\,.
\end{equation}
(Note that this is not a path integral; all the fields in \eqref{sigmadef} are functions of space at a fixed time.) The continuity of the field demands that $\phi_D$ and $\phi_{D^c}$ agree on the mutual boundary $\partial D$, and similarly for $\phi_D'$ and $\phi_{D^c}$; hence $\phi_D$ and $\phi_D'$ must agree. In other words $\sigma$ is block-diagonal in the field value on $\partial D$,\footnote{This statement is strictly true only in the limit where the UV cutoff is removed. In the presence of a cutoff that produces a well-defined quantum-mechanical system, such as a lattice, the eigenvectors of $\sigma$ must be normalizable, and hence cannot also be eigenvectors of $\hat\phi(x)$, which has a continuous spectrum. Indeed, the presence of a cutoff such as a lattice relaxes the continuity requirement: With $x_1$ and $x_2$ being neighboring lattice points inside and outside $D$ respectively, the wave functional $\ev{\phi_D,\phi_{D^c}|0}$ is merely sharply peaked, but not a delta-function, about configurations where $\phi(x_1)=\phi(x_2)$; hence the matrix elements of $\sigma$ do not strictly vanish but are merely suppressed when $\phi(x_1)\neq\phi'(x_1)$.}
\begin{equation}
\ev{\phi_D|\sigma|\phi_D'} \propto \delta\left[\left.\phi_D\right|_{\partial D}-\left.\phi_D'\right|_{\partial D}\right].
\end{equation}

We can also use the fact that this theory is a CFT\footnote{More precisely, it can be made conformal by adding to the Lagrangian a conformal mass term, which vanishes in flat space.}  to express $\sigma$ in terms of the generator $K$ of the conformal transformation on Euclidean $\R^3$ (or Minkowski space) that fixes the edge of the disk: $\sigma = e^{-K}/(\Tr e^{-K})$.
Near the edge of the disk, this conformal transformation looks
like a rotation (or boost); in particular, the Weyl scaling factor is 1 on the edge of the disk.
It follows that $K$, and therefore $\sigma$, commutes with $\hat\phi(x)$, since that is a scalar primary operator, for $x\in\partial D$.

As an aside, we note that these two proofs admit a few interesting immediate generalizations. The first proof applies for any dimensionality of spacetime, any entangling surface, and any state. The second proof applies to any scalar primary operator in any CFT in any dimension. While it only applies to a spherical entangling surface (since only then can the reduced density matrix be written in terms of a conformal generator), it seems reasonable to expect that, being purely local, the property will hold for a general entangling surface.

The fact that the field has a definite value on the entangling surface for each pure state in $\sigma$ can be understood heuristically in a several ways. First, it reflects the fact that entanglement leads to decoherence: the field values at neighboring points are highly entangled, so if we trace over one of them, then the field value at the other point will become decohered. Another point of view is to regard $\sigma$ as a mixed state of the scalar
theory quantized on $D$, without reference to the rest of the
system. Consistently quantizing the theory on $D$ requires
specifying some boundary conditions on $\partial D$, such as
Dirichlet boundary conditions in which $\phi|_{\partial
D}=\phi_0(x)$ for some given function $\phi_0(x)$ on $\partial D$.
Different choices of $\phi_0(x)$ define different theories, or, to put it another way, different superselection sectors. Hence the
density matrix $\sigma$ can contain mixtures but not superpositions of pure
states with different values of $\phi|_{\partial D}$.

\subsection{Compact scalar}\label{sec:compactscalar}

We now turn to the compact theory. This can be regarded as a $\Z$ gauging of the non-compact theory, where the gauge transformations act as global discrete shifts $\phi\to\phi+m$ ($m\in\Z$). On the plane, spontaneous symmetry breaking implies that $\phi$ must go to a fixed constant at infinity, which without loss of generality we can take to be 0 mod 1; by gauge fixing, we can simply take that value to be 0. Since $\R^2$ is topologically trivial, the set of field configurations in this gauge is the same as in the non-compact theory. It is easy to see using either canonical quantization or the Euclidean path integral that the wave functional $\ev{\phi|0}$ is also the same (in this gauge) as in the non-compact theory.\footnote{In the canonical quantization, the zero-mode is fixed to be zero just as in the non-compact theory, while the non-zero-modes are unaffected by the gauging. In the Euclidean path integral, the wave functional $\ev{\phi|0}$ is given by the path integral on half of $\R^3$ with Dirichlet boundary conditions (the field goes to $\phi$ on the boundary $\R^2$ and, in this gauge, to 0 at infinity); since half of $\R^3$ is also topologically trivial, the set of three-dimensional field configurations entering into the path integral is the same as in the non-compact theory, as is their action.}

Now that we have the wave functional for the compact theory, we wish to construct the reduced density matrix $\rho$. A field configuration on $D$ in the compact theory can be regarded as an equivalence class under constant integral shifts of field configurations in the non-compact theory. If we gauge fix in some manner, so that each compact field configuration is represented by a single non-compact one $\phi_D$, then the continuity requirement across $\partial D$ is relaxed; rather than $\phi_D|_{\partial D}=\phi_{D^c}|_{\partial D}$ as in the non-compact theory, $\phi_D|_{\partial D}-\phi_{D^c}|_{\partial D}$ is only required to be a constant integer. (We continue to gauge-fix $\phi_{D^c}$ as before, i.e.\ to require that it go to 0 at infinity.) So for the wave functional we have
\begin{equation}
\ev{\phi_D,\phi_{D^c}|0}_{\rm compact} = \sum_m\ev{\phi_D+m,\phi_{D^c}|0}_{\rm non-compact}\,,
\end{equation}
and hence
\begin{equation}\label{rhodef1}
\ev{\phi_D|\rho|\phi_D'} = \sum_{m,m'}\ev{\phi_D+m|\sigma|\phi'_D+m'}\,.
\end{equation}
Using a convenient choice of gauge and the fact that $\sigma$ is block-diagonal, the double sum can be reduced to a single sum. For example, pick an arbitrary point $x_0\in\partial D$, and require $0\le\phi_D(x_0)<1$. Then only the terms with $m=m'$ contribute to the sum in \eqref{rhodef1}. Defining the operator $\rho_m$ on the compact Hilbert space with matrix elements
\begin{equation}
\ev{\phi_D|\rho_m|\phi_D'} = \ev{\phi_D+m|\sigma|\phi'_D+m}\,,
\end{equation}
we have
\begin{equation}\label{rho}
\rho=\sum_m\rho_m\,.
\end{equation}

As a check on \eqref{rho}, we will show that it leads to the same path integral for calculating the R\'enyis as was used in subsection \ref{sec:Renyis}.\footnote{The fact that the replica trick correctly reproduces the R\'enyi entropies in the presence of a discrete gauge symmetry was shown by a similar construction in \cite{Headrick:2012fk}.} First, the matrix element $\ev{\phi_D|\rho_m|\phi_D'}$ is given by a Euclidean path integral on $\R^3\setminus D$ with Dirichlet boundary conditions: $\phi$ asymptotes to 0 at infinity, and equals $\phi_D+m$ and $\phi'_D+m$ respectively on the ``top'' and ``bottom'' of $D$. Equivalently, by invariance under constant shifts, this is the path integral where $\phi$ asymptotes to $-m$ at infinity and equals $\phi_D$ and $\phi_D'$ respectively on the top and bottom of $D$. We then have
\begin{equation}
\exp\left((1-n)S_n(r)\right) = \Tr\rho^n = \sum_{m_1,\ldots,m_n}\Tr\rho_{m_1}\cdots\rho_{m_n}\,.
\end{equation}
More explicitly,
\begin{equation}\label{gfpathintegral}
\Tr\rho_{m_1}\cdots\rho_{m_n} =
\int_{0\le\phi(x_0)<2\pi R}[d\phi_{D1}d\phi_{Dn}]\,\ev{\phi_{D1}|\rho_{m_1}|\phi_{D2}}\ldots\ev{\phi_{Dn}|\rho_{m_n}|\phi_{D1}}\,.
\end{equation}
This is given by a path integral on the $n$-fold branched cover $E_n(r)$ defined in subsection \ref{sec:Renyis}, with the field asymptoting to $m_i$ on the $i$th sheet. Note that, since $x_0$ lies on the branch locus $\partial D$, $\phi(x_0)$ has a single value, so the gauge condition $0\le\phi(x_0)<g$ only imposes a single restriction on the integral. To remove this restriction, we shift the field globally by $m_n$ and sum over $m_n$. The field then asymptotes to $m_n-m_i$ on the $i$th sheet. Setting $w_i=m_n-m_i$ ($i=1,\ldots,n-1$), we have a sum over $w$ of unrestricted path integrals on $E_n(r)$, where the field asymptotes to $w_i$ on the $i$th sheet and to 0 on the $n$th sheet. This is precisely the path integral used to compute $S_n(r)$ in subsection \ref{sec:Renyis}.

\subsection{Relation between the two density matrices}

We can now reconstruct $\sigma$ from the operators $\rho_m$. If we decompose the non-compact Hilbert space according to the value of $\phi_D(x_0)$,
\begin{equation}
\mathcal{H}_{\rm non-compact} = \bigoplus_m\mathcal{H}_m\,,\qquad
\mathcal{H}_m = \Hspan\{\ket{\phi_D}:mg\le\phi_D(x_0)<(m+1)g\}\,,
\end{equation}
then there is a natural isomorphism $U_m$ from $\mathcal{H}_m$ to $\mathcal{H}_{\rm compact}$: $U_m\ket{\phi_D} = \ket{\phi_D-mg}$. Then clearly
\begin{equation}
\sigma=\bigoplus_mU_m^{-1}\rho_mU_m\,.
\end{equation}

It is easy to show that $\rho$ is a purer state than $\sigma$, meaning that the spectrum of $\rho$ majorizes that of $\sigma$. First we show that, for any positive numbers $p_i$ ($p_1\ge p_2\ge\cdots$) and normalized (not necessarily orthogonal) states $\ket{i}$, the spectrum of $\rho=\sum p_i\ket{i}\bra{i}$ majorizes the set $\{p_i\}$: if $P_k$ is the orthogonal projector onto the space spanned by $\{\ket{i}:i=1,\ldots,k\}$ and $\lambda_i$ are the eigenvalues of $\rho$ in decreasing order, then
\begin{equation}
\sum_{i=1}^kp_i=\Tr\rho P_k\le\sum_{i=1}^k\lambda_i\,.
\end{equation}
Now let the eigenvectors and eigenvalues of $\rho_m$ be $\ket{mj}$, $p_{mj}$ respectively. Then the spectrum of $\sigma$ is simply the collection $\{p_{mj}\}$ over all $m,j$, while $\rho=\sum_{m,j}p_{mj}\ket{mj}\bra{mj}$, so we can apply the previous statement. Finally, it is well known that a purer state has a smaller R\'enyi entropy for any $n$ (including $n=1$). (See Section II of \cite{Wehrl} for a complete discussion of relative purity and R\'enyi entropies.)

The reasoning of this section also explains to some extent why the R\'enyis are increasing functions of $g$, and hence of $r$ (see figure \ref{fig:Splot}, and recall that $r$ there is really $rg^2$). For any positive integer $k$, the theory with periodicity $g$ can be regarded as a $\Z_k$ gauging of the theory with periodicity $kg$, and therefore has smaller R\'enyis.

\acknowledgments
We benefitted from helpful discussions with Horacio Casini, Igor Klebanov, Zohar Komargodski, Albion Lawrence, Shiraz Minwalla, Silviu Pufu, Matthew Roberts, Howard Schnitzer, Erik Tonni, and Laurence Yaffe. We would also like to thank Horacio Casini, Igor Klebanov, Zohar Komargodski, Albion Lawrence, Silviu Pufu, and Erik Tonni for helpful comments on a draft of this paper. M.H. would like to thank the Tata Institute of Fundamental Research for hospitality while some of this work was completed. The work of M.H. was supported in part by the National Science Foundation under CAREER Grant No.\ PHY10-53842. 
The research of D.J. was supported by the Department of Energy Award \# DE-SC0007870, and the Fundamental Laws Initiative of the Center for the Fundamental Laws of Nature, Harvard University.

\appendix

\section{Three-sphere free energy}\label{freeenergy}

Here we calculate the renormalized $S^3$ free energy of the periodic scalar theory, using a zeta-function regularization.\footnote{This result was independently obtained by S. Pufu (private communication).} This theory cannot be conformally coupled on the sphere, since a conformal mass term would be inconsistent with the periodicity. Thus this partition function misses the spontaneous symmetry breaking that plays an important role in the entanglement entropy. It does, however, nicely match the logarithmic behavior in the UV, as discussed in subsection \ref{sec:results}.

The sphere has radius $r$ and the scalar has period $g$. Let the eigenvalues of the scalar Laplacian on the unit $S^3$ be $\lambda_k$; then the eigenvalues on the sphere of radius $r$ are $\lambda_k/r^2$. We normalize the eigenfunctions such that their average value on the sphere is 1:
\begin{equation}
\frac1{2\pi^2r^3}\int\phi_k\phi_{k'}=\delta_{kk'}\,.
\end{equation}
 We will not need the precise form of the eigenvalues and eigenfunctions, except that $\lambda_0=0$ and $\phi_0=1$. A general field configuration is $\phi = \sum c_k\phi_k$, where $c_0$ is periodic with period $g$ while the other $c_k$ run over the real line. The action for this configuration is
\begin{equation}
S[\phi]=\pi^2 r\sum_k\lambda_kc_k^2\,.
\end{equation}

The partition function is
\begin{equation}
Z = \int[d\phi]e^{-S[\phi]}
=\int_0^g\frac{dc_0}{\mu_0^{1/2}}\prod_{k\neq0}\int\frac{dc_k}{\mu_0^{1/2}}e^{-\pi^2r\lambda_kc_k^2}\,;
\end{equation}
$\mu_0$ is an arbitrary renormalization scale that drops out in zeta-function regularization, reflecting the absence of a Weyl anomaly in three dimensions (see for example \cite{Klebanov:2011gs}). For convenience we set $\mu_0=1/(\pi r)$. We then obtain
\begin{eqnarray}\label{free2}
F_{S^3} &:=&-\ln Z\nonumber \\ 
&=& -\frac12\ln(\pi rg^2) +\frac12\sum_{k\neq0}\ln\lambda_k \nonumber\\
&= & -\frac12\ln(\pi rg^2) +\frac12\ln\det{}'(-\nabla^2_1) \nonumber\\
&=&-\frac12\ln(rg^2)+\frac{\zeta(3)}{4\pi^2}\,;
\end{eqnarray}
here $\det'(-\nabla^2_1)$ is the determinant over non-zero modes of the Laplacian on the unit sphere, whose value can be found, for example, in \cite{Dowker:1993jv,Klebanov:2011td}. \eqref{free2} agrees with the result obtained in \cite{Klebanov:2011td} using the Maxwell description of the theory.

\section{Determination of $J(\beta)$}\label{Jcalc}

Our task in this appendix---required for the calculation of the R\'enyi entropies in subsection \ref{sec:Renyis}---is to find the complex solution $\tilde\phi_\beta$ to Laplace's equation in three dimensions that asymptotes to 1 at infinity and has a phase monodromy of $e^{2\pi i\beta}$ around the unit circle (on the $\tau=0$ plane), and to calculate its classical action
\begin{equation}
J(\beta)\equiv S[\tilde\phi_\beta] = \frac12\int d^3\!y\,\partial_\mu\tilde\phi_\beta^*\partial^\mu\tilde\phi_\beta\,.
\end{equation}
We will hereafter simply refer to $\tilde\phi_\beta$ as $\phi$. Actually, we will not completely succeed in this task, in the sense that we will not be able to provide an explicit solution. Instead, we will give what we believe is convincing evidence that the function $J(\beta)$ is given by \eqref{Jbeta}, which we repeat here for convenience:
\begin{equation}\label{Jbeta2}
J(\beta) = 2\pi(1-2\beta)\tan(\pi\beta)
\end{equation}
(where, as in subsection \ref{sec:Renyis}, $\beta$ is taken to lie between 0 and 1).

This appendix is organized as follows. In subsection \ref{actionflux}, we will show that the on-shell classical action $J(\beta)$ is equal to the coefficient of the leading fall-off of the field at infinity. In subsection \ref{generalsolution}, working in an oblate spheroidal coordinate system, we will write the general solution to Laplace's equation away from the disk as a linear combination of Legendre functions, with $J(\beta)$ equal to the leading coefficient. In subsection \ref{constraint}, we will write the monodromy condition as a matrix equation for the coefficients. Finally, in subsection \ref{evidence}, we will show that three independent approximation schemes for the matrix equation yield results that are consistent with \eqref{Jbeta2}.

\subsection{Action as flux}\label{actionflux}

We begin in the usual cylindrical coordinates $(\tau,\rho,\varphi)$,
\begin{equation}\label{bc1}
ds^2 = d\tau^2+d\rho^2+\rho^2d\varphi^2\,,
\end{equation}
where we place the circle at $\rho=1$ on the $\tau=0$ plane. (The coordinate $\tau$, which originates as the Euclidean time direction in the replica trick, is usually called $z$ in cylindrical coordinates.) We demand that $\phi$ is subject to a monodromy of $e^{2\pi i\beta}$ upon going around the unit circle $\tau=0$, $\rho=1$. Placing the branch cut on the unit disk $D=\{(\tau,\rho,\phi):\tau=0,\rho<1\}$, we require
\begin{equation}\label{bc2}
\phi(\tau=0^+,\rho,\varphi) = e^{2\pi i\beta}\phi(\tau=0^-,\rho,\varphi)\,.
\end{equation}
Since we have a second-order equation of motion, we need a boundary condition on the first derivative of the field as well; this is given by continuity of the first derivative up to multiplication by $e^{2\pi i\beta}$:
\begin{equation}\label{bc3}
\partial_\tau\phi(\tau=0^+,\rho,\varphi) = e^{2\pi i\beta}\partial_\tau\phi(\tau=0^-,\rho,\varphi)\,.
\end{equation}

Let us make a comment about the behavior of $\phi$ on the unit circle. By the monodromy condition \eqref{bc2}, it necessarily goes either to 0 or infinity there. We are interested in the (unique) solution in which it goes to 0. That solution will have finite action, which can be seen as follows. Very close to the circle, we can ignore the $\varphi$ component of the metric, and the solution to Laplace's equation will be approximately harmonic in the $\tau,\rho$ plane. Therefore it will be of the form $f(z)+g(z)^*$, where $z=\rho-1+i\tau$ and $f$ and $g$ are holomorphic with branch cuts along the negative real axis. The monodromy condition requires $f(z)\sim z^\beta$, $g(z)\sim z^{1-\beta}$. The dominant behavior near $z=0$ will thus be $z^\beta$ for $0<\beta<1/2$ and $\bar z^{1-\beta}$ for $1/2<\beta<1$, and in both cases the action is integrable.

Now we note that, when $\phi$ solves the equation of motion, the action reduces to a surface integral:
\begin{equation}\label{surface}
S[\phi] = \frac12\int_\partial d^2\!y\,h^{1/2}n^\mu\phi^*\partial_\mu\phi\,,
\end{equation}
where $h$ is the determinant of the induced metric on the surface and $n^\mu$ is an outward-directed normal vector. This surface includes a sphere at infinity. Since the field does not actually solve the equation of motion on the disk $D$, where its derivative is discontinuous, we must exclude $D$ from the region of integration, leading to another surface on which we must evaluate \eqref{surface}, namely a surface enclosing $D$. However, the boundary conditions \eqref{bc1}, \eqref{bc2} imply that \eqref{surface} actually vanishes on the latter surface, leaving only the contribution from the sphere at infinity. Since $\phi$ goes to 1 at infinity, the action simply equals (half) the flux at infinity:
\begin{equation}
S[\phi] = \frac12\int_\infty d^2\!y\,h^{1/2}n^\mu\partial_\mu\phi\,.
\end{equation}
By the equation of motion, this in turn equals the flux through any sphere enclosing $D$. Due to the discontinuity in the first derivative \eqref{bc3}, there is effectively a source for the field at $D$.

\subsection{General solution}\label{generalsolution}

To solve the equation of motion, it is convenient to change to oblate spheroidal coordinates, which are better adapted to this geometry. This is an orthogonal coordinate system consisting of a radial coordinate $\zeta$ ($\zeta\ge0$), a polar coordinate $\eta$ ($-1\le\eta\le1$; $\eta$ is analogous to $\cos\theta$ in spherical coordinates), and the usual azimuthal coordinate $\varphi$. The relation to cylindrical coordinates is
\begin{equation}
\tau = \zeta\eta\,,\qquad\rho = \sqrt{(1+\zeta^2)(1-\eta^2)}\,.
\end{equation}
The metric is
\begin{equation}\label{metric}
ds^2 = \frac{\zeta^2+\eta^2}{1+\zeta^2}d\zeta^2+\frac{\zeta^2+\eta^2}{1-\eta^2}d\eta^2+(1+\zeta^2)(1-\eta^2)d\varphi^2\,.
\end{equation}
The surfaces of constant $\zeta$ are oblate ellipsoids, with $\zeta=0$ being the unit disk, while the surfaces of constant $\eta$ are hyperboloids, with $\eta=\pm1$ being the $\tau$-axis and $\eta=0$ being the $\tau=0$ plane excluding the unit disk. The unit circle is at $\zeta=\eta=0$. The Laplacian is
\begin{equation}
\nabla^2 = \frac1{\zeta^2+\eta^2}\left(\partial_\zeta(1+\zeta^2)\partial_\zeta+\partial_\eta(1-\eta^2)\partial_\eta+\partial_\varphi\frac{\zeta^2+\eta^2}{(1+\zeta^2)(1-\eta^2)}\partial_\varphi\right).
\end{equation}
In this coordinate system the boundary conditions on $\phi(\zeta,\eta,\varphi)$ are at $\zeta=0,\infty$:
\begin{equation}
\lim_{\zeta\to\infty}\phi(\zeta,\eta,\varphi) = 1\,,
\end{equation}
\begin{equation}\label{ellbc}
\left.
\begin{array}{c}\phi(0,\eta,\varphi) = e^{2\pi\beta i}\phi(0,-\eta,\varphi)\\
\partial_\zeta\phi(0,\eta,\varphi) = -e^{2\pi\beta i}\partial_\zeta\phi(0,-\eta,\varphi)
\end{array}\right\}\quad\eta>0\,.
\end{equation}

The equation of motion and the boundary conditions are invariant under rotations about the $\tau$-axis, so we can assume that the solution is, as well. From now on, therefore, we will drop any dependence on the $\varphi$ coordinate. The equation and boundary conditions are also invariant under a combined reflection through the $\tau=0$ plane ($\eta\to-\eta$) and complex conjugation of $\phi$. Since the equation is linear, the solution must share this symmetry, i.e.\ $\phi(\zeta,\eta) = \phi^*(\zeta,-\eta)$.

Under separation of variables, Laplace's equation yields Legendre's equation in both $\zeta$ and $\eta$:
\begin{eqnarray}
\left(\partial_\zeta^2+2\frac\zeta{1+\zeta^2}\partial_\zeta-\frac{l(l+1)}{1+\zeta^2}\right)Z(\zeta) &= 0 \\
\left(\partial_\eta^2-2\frac\eta{1-\eta^2}\partial_\eta+\frac{l(l+1)}{1-\eta^2}\right)H(\eta) &= 0
\end{eqnarray}
As usual with spherical coordinates, the solutions to the $\eta$ equation, requiring regularity at $\eta=\pm1$, are Legendre polynomials $P_l(\eta)$, $l=0,1,2,\ldots$. The solutions to the $\zeta$ equation are $P_l(i\zeta),Q_l(i\zeta)$, where $Q_l$ are Legendre functions of the second kind.\footnote{Not all references agree on the definition of the Legendre function of the second kind. Here by $Q_l(i\zeta)$ we mean the function obtained in \emph{Mathematica} as \texttt{LegendreQ[l, 0, 3, I zeta]}, or equivalently \texttt{LegendreQ[l, I zeta] - I Pi/2 LegendreP[l, I zeta]}. For example, $Q_0(i\zeta) = \frac12\ln((1+i\zeta)/(1-i\zeta))=-i\arccot\zeta$, $Q_1(i\zeta) = \zeta\arccot\zeta-1$.} However, $P_l(i\zeta)$ blows up as $\zeta\to\infty$ for $l>0$, while $Q_l(i\zeta)$ goes to 0 (like $\zeta^{-(l+1)}$) for all $l$. Therefore for $l>0$ we can only include $Q_l(i\zeta)$, while for $l=0$ we can include both solutions. Since $P_0(i\zeta)=P_0(\eta)=1$, we must include the $P_0$ solution with unit coefficient in order to satisfy our boundary condition at infinity. The general solution satisfying the boundary condition at infinity can therefore be written
\begin{equation}\label{gensol}
\phi(\zeta,\eta) = 1-i\sum_{l=0}^\infty c_lQ_l(i\zeta)P_l(\eta)\,.
\end{equation}
A factor of $i$ has been inserted in order to make the $c_l$ coefficients real, given the symmetry under reflection and complex conjugation. (Note that $P_l(\zeta)$ has the same parity as $l$, while $Q_l(i\zeta)$ is imaginary (real) for even (odd) values of $l$.)

In calculating the flux through any ellipsoid of constant $\zeta$, the only contribution is from the $l=0$ term in \eqref{gensol}, since $P_l(\eta)$ integrates to 0 for $l>0$. Since, for large $\zeta$, $Q_0(i\zeta)=-i\zeta^{-1}+\mathcal{O}(\zeta^{-3})$, we have
\begin{equation}
J(\beta) = 2\pi c_0\,.
\end{equation}
Our task is thus to find $c_0$.

\subsection{Monodromy conditions and formal solution}\label{constraint}

The only constraints that are not built into the general solution \eqref{gensol} are the monodromy conditions \eqref{ellbc} on the disk. These conditions are not very simple to impose since they are non-local in $\eta$. In this subsection we will convert them into a matrix equation for the vector of coefficients $c_l$, which can be formally solved.

We begin by noting that the conditions \eqref{ellbc} have definite parity under $\eta \to -\eta $, that is, $e^{-\pi\beta i\sgn\eta}\phi(0,\eta)$ is even while $e^{-\pi\beta i\sgn\eta}\partial_\zeta\phi(0,\eta)$ is odd. It is thus useful to split $\phi$ into its real and imaginary parts, which are even and odd in $\eta$ respectively:
\begin{equation}
\phi = \phi_++i\phi_-\,,\qquad\phi_+(\zeta,\eta) = 1-i\sum_{l\text{ even}}c_lQ_l(i\zeta)P_l(\eta)\,,\qquad\phi_-(\zeta,\eta)=-\sum_{l\text{ odd}}c_lQ_l(i\zeta)P_l(\eta)\,.
\end{equation}
In terms of $\phi_\pm$, \eqref{ellbc} becomes
\begin{eqnarray}\label{bcphipm}
\phi_+(0,\eta)&=&\lambda\sgn\eta\,\phi_-(0,\eta) \label{bcphipm1} \\
\partial_\zeta\phi_-(0,\eta)&=&-\lambda\sgn\eta\,\partial_\zeta\phi_+(0,\eta) \,, \label{bcphipm2}
\end{eqnarray}
where
\begin{equation}
\lambda:=\cot(\pi\beta)
\end{equation}
(note that $\lambda$ ranges from $\infty$ to $-\infty$ as $\beta$ goes from 0 to 1).

We now multiply \eqref{bcphipm1}, \eqref{bcphipm2} by $P_{l'}(\eta)$ and integrate over $\eta$. Since these two equations are even and odd, respectively, it suffices to take $l'$ even and odd. From \eqref{bcphipm1} we obtain
\begin{equation}\label{even}
\sqrt2\delta_{l'0} - i\frac{c_{l'}Q_{l'}(0)}{\sqrt{l'+\frac12}} = -\lambda\sum_{l\text{ odd}}S_{l'l}\frac{c_lQ_l(0)}{\sqrt{l+\frac12}}\qquad(l'\text{ even})\,,
\end{equation}
where
\begin{equation}
S_{l'l} := \sqrt{(l+\frac12)(l'+\frac12)}\int_{-1}^1d\eta\,\sgn\eta P_l(\eta)P_{l'}(\eta)\,.
\end{equation}
(The square-root factors in the definition of $S$ are for later convenience.) Similarly, \eqref{bcphipm2} becomes
\begin{equation}\label{odd}
\frac{c_{l'}Q'_{l'}(0)}{\sqrt{l'+\frac12}} = -i\lambda\sum_{l\text{ even}}\frac{c_lQ'_l(0)}{\sqrt{l+\frac12}}S_{ll'}\qquad (l'\text{ odd})\,.
\end{equation}
Both \eqref{even} and \eqref{odd} are linear equations connecting the even and odd coefficients $c_l$. The former is inhomogeneous while the latter is homogeneous. Combining them to eliminate the odd coefficients, we obtain:
\begin{equation}\label{overall}
\sqrt2\delta_{l0} - i\frac{c_lQ_l(0)}{\sqrt{l+\frac12}} = i\lambda^2\sum_{\substack{l'\text{ odd}\\l''\text{ even}}}S_{ll'}\frac{Q_{l'}(0)}{Q_{l'}'(0)}S_{l''l'}\frac{c_{l''}Q'_{l''}(0)}{\sqrt{l''+\frac12}}\qquad(l\text{ even})\,.
\end{equation}
This inhomogeneous linear equation in principle admits a unique solution.

For clarity, it is useful to recast \eqref{overall} in matrix form. We define the vectors
\begin{equation}
d = \left(\frac{c_lQ_l'(0)}{\sqrt{l+\frac12}}\right)_{l\text{ even}},\qquad
u = \left(\delta_{l0}\right)_{l\text{ even}},
\end{equation}
and the matrices
\begin{equation}
Q_+ = \diag\left(i\frac{Q_l(0)}{Q_l'(0)}\right)_{l\text{ even}},\qquad
Q_- = \diag\left(i\frac{Q_l(0)}{Q_l'(0)}\right)_{l\text{ odd}},\qquad
S = \left(S_{ll'}\right)_{l\text{ even, }l'\text{ odd}}\,.
\end{equation}
Explicit expressions for these vectors and matrices will be given shortly (except, of course, for the values of the $c_l$ appearing in $d$). For now, note that they are real, and that (by the orthogonality and completeness of the Legendre polynomials, together with the fact that $(\sgn\eta)^2=1$) $S$ is orthogonal: $SS^T=S^TS=I$. \eqref{overall} can now be written
\begin{equation}
\sqrt2u - Q_+d = \lambda^2 S Q_-S^Td\,,
\end{equation}
whose solution is
\begin{equation}
d = \sqrt2 Q_+^{-1}\left(I+\lambda^2 T\right)^{-1}u\,,
\end{equation}
where
\begin{equation}\label{Tdef}
T:=SQ_-S^TQ_+^{-1}\,.
\end{equation}
Using the fact that $J(\beta)=2\pi c_0=\sqrt2\pi u^Td$ and $u^TQ_+^{-1}= 2 u^T/\pi$ (see below), we have
\begin{equation}\label{c0final}
J(\beta) =4 u^T\left(I+\lambda^2 T\right)^{-1}u\,.
\end{equation}
This is our formal solution: $J(\beta)$ is 4 times the $00$ component of the inverse of the infinite-dimensional matrix $I+\lambda^2 T$.

Let us now record explicit expressions for the components of the vectors and matrices defined in the last paragraph. From section 14.5 of \cite{NIST}, for even $l$ we have
\begin{equation}
Q_l(0) = -i\frac\pi2\frac{l!}{(-4)^{l/2}(l/2)!^2}\,, \qquad
Q_l'(0) = \frac{(-4)^{l/2}(l/2)!^2}{l!}\,,
\end{equation}
so
\begin{equation}\label{Qplus}
(Q_+)_{ll} = \frac\pi2\frac{l!^2}{4^l(\frac l2!)^4}\,,
\end{equation}
and for odd $l$,
\begin{equation}
Q_l(0) = -\frac{(-4)^{(l-1)/2}(\frac{l-1}2)!^2}{l!}\,,\qquad
Q'_l(0) = -i\frac\pi2\frac{l!}{(-4)^{(l-1)/2}(\frac{l-1}2)!^2}
\end{equation}
so
\begin{equation}\label{Qminus}
(Q_-)_{ll} = \frac2\pi\frac{4^{l-1}(\frac{l-1}2!)^4}{l!^2}\,.
\end{equation}
We can also derive a closed-form expression for the components of $S$,
using a procedure analogous to the derivation of the orthogonality property of the Legendre polynomials.\footnote{In the process we multiply the Legendre differential equation by an arbitrary Legendre polynomial and subtract from the result an identical expression with the two Legendre polynomials' subscripts reversed. This allows us to rewrite the integral of a product of Legendre polynomials as an integral of total derivatives, which is trivially computed. For an explicit derivation, see chapter 5 of \cite{BWE}.} The resulting expression is:
\begin{equation}\label{Smatrix}
 S_{ll'}=\frac{(-1)^{(l+l'+1)/2}\sqrt{l+\frac 12}\sqrt{l'+\frac 12}l!l'!}{2^{l+l'-2}(l-l')(l+l'+1)(\frac{l}{2}!)^2(\frac{l'-1}{2}!)^2}\qquad \text{($l$ even, $l'$ odd)}\,.
\end{equation}
From the definition of the $T$ matrix, its elements are given by an infinite sum that can be rewritten in terms of the digamma and trigamma functions $\psi$, $\psi_1$ respectively (the first and second derivatives of $\ln\Gamma$) using their series representations \cite{SC}:
 \begin{equation}\label{Tfunction}
           T_{ll'}= \begin{cases}\displaystyle\frac{2(-1)^{(l+l')/2}(2l+1)\frac{l'}{2}!\frac{l-1}{2}![\psi(\frac{1-l}{2}) + \psi(1+\frac l2) -\psi(\frac{1-l'}{2}) -\psi(1 + \frac{ l'}{2})]}{\pi^2 \frac l2!\frac{l'-1}{2}!(l'-l)(l+l'+1)}    &\quad(l'\neq l)\\
\displaystyle                 \frac{\psi_1(\frac{1-l}2)-\psi_1(1+\frac l2)}{\pi^2}&\quad(l'=l)
                 \end{cases}
       \end{equation}
(where $l,l'$ are even).

\subsection{Evidence for the functional form of $J(\beta)$\label{evidence}}

Despite having the matrix $T$ in closed form, we were not able to calculate analytically the inverse of $I+\lambda^2T$, or even its 00 element. We will therefore present three sets of calculations providing independent evidence that the 00 element of the inverse is given as follows:
\begin{equation}\label{Jfinal}
u^T(I+\lambda^2T)^{-1}u = \frac{\arctan\lambda}\lambda\,;
\end{equation}
using \eqref{c0final}, \eqref{Jfinal} implies \eqref{Jbeta2}.\footnote{We were also able to guess and confirm the following functional forms for $c_1$ and $c_2$:
\begin{eqnarray}
c_1 &=& \frac{6}{\pi^2}\frac{\arctan^2\lambda}{\lambda} = \frac{3}{2}(1-2\beta)^2\tan(\pi\beta)\\
c_2&=&\frac{10\sqrt{5}}{\pi^3}\frac{\arctan^3\lambda}{\lambda} = \frac{5\sqrt{5}}{4}(1-2\beta)^3\tan(\pi\beta)\,.
\end{eqnarray}
However, we were not able to find the form of $c_l$ for general $l$.}

\begin{figure}[tbp]
\centering
\includegraphics[width=.65\textwidth]{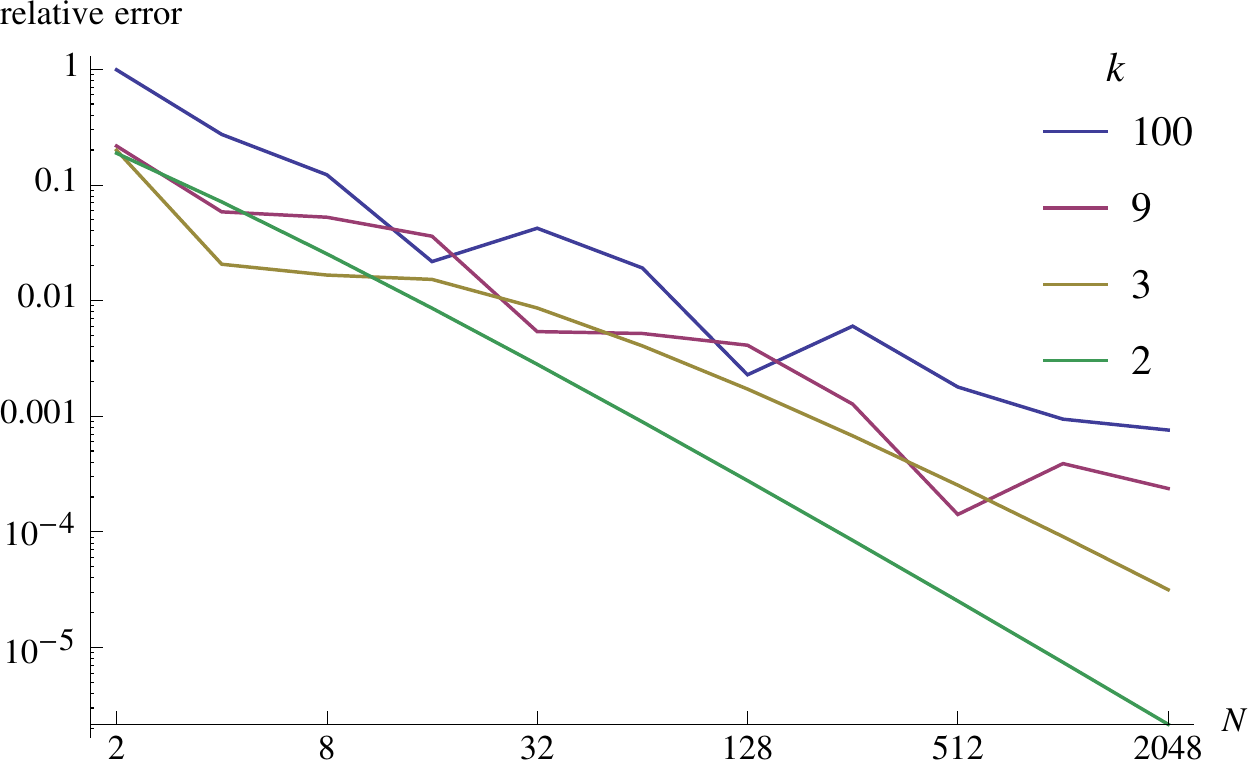}
\caption{\label{fig:Tk00error} Relative error in \eqref{match} when $T$ is truncated to an $N\times N$ matrix, for a few sample values of $k$.}
\end{figure}

\paragraph{Expansion about $\lambda=0$:} The function $\lambda^{-1}\arctan\lambda$ is analytic at $\lambda=0$, with Taylor series
\begin{equation}\label{sum}
\sum_{k=0}^\infty\frac{(-1)^k}{2k+1}\lambda^{2k}\,.
\end{equation}
We can similarly expand $(I+\lambda^2T)^{-1}$ in powers of $\lambda$. If \eqref{Jfinal} is correct, then we should find
\begin{equation}\label{match}
u^TT^ku = \frac1{2k+1}\,.
\end{equation}
This clearly holds for $k=0,1$. For $k>1$, numerical evaluation of the right-hand side using truncated matrices confirmed \eqref{match} to high precision and with good convergence as the size of the matrices is increased; see figure \ref{fig:Tk00error} for a few examples.

\paragraph{Expansion about $\lambda=\infty$:} We can also expand $(I+\lambda^2T)^{-1}$ in $\lambda^{-1}$:
\begin{equation}
u^T(I+\lambda^2T)^{-1}u = u^TT^{-1}u\,\lambda^{-2}+\mathcal{O}(\lambda^{-4})\,.
\end{equation}
To calculate the matrix element of $T^{-1}$ we can refer back to the definition \eqref{Tdef}:
\begin{equation}
T^{-1} = Q_+SQ_-^{-1}S^T
\end{equation}
(where we used the fact that $S$ is orthogonal); using \eqref{Qplus}, \eqref{Qminus}, \eqref{Smatrix}, we obtain
\begin{equation}
u^TT^{-1}u = \frac12\sum_{l\text{ odd}}\frac{l^2(l+\frac12)\Gamma(\frac l2)^4}{(l+1)^2\Gamma(\frac{l+1}2)^4}\,.
\end{equation}
However, the summand goes like $l^{-1}$ for large $l$, so the sum diverges logarithmically. Hence $u^T(I+\lambda^2T)^{-1}u$ is non-analytic at $\lambda^{-1}=0$; specifically, it vanishes but has infinite second derivative. Indeed, this is precisely the behavior of $\lambda^{-1}\arctan\lambda$, which for $\lambda^{-1}\approx0$ goes like $|\lambda^{-1}|$.

\begin{figure}[tbp]
\centering
\includegraphics[width=.95\textwidth]{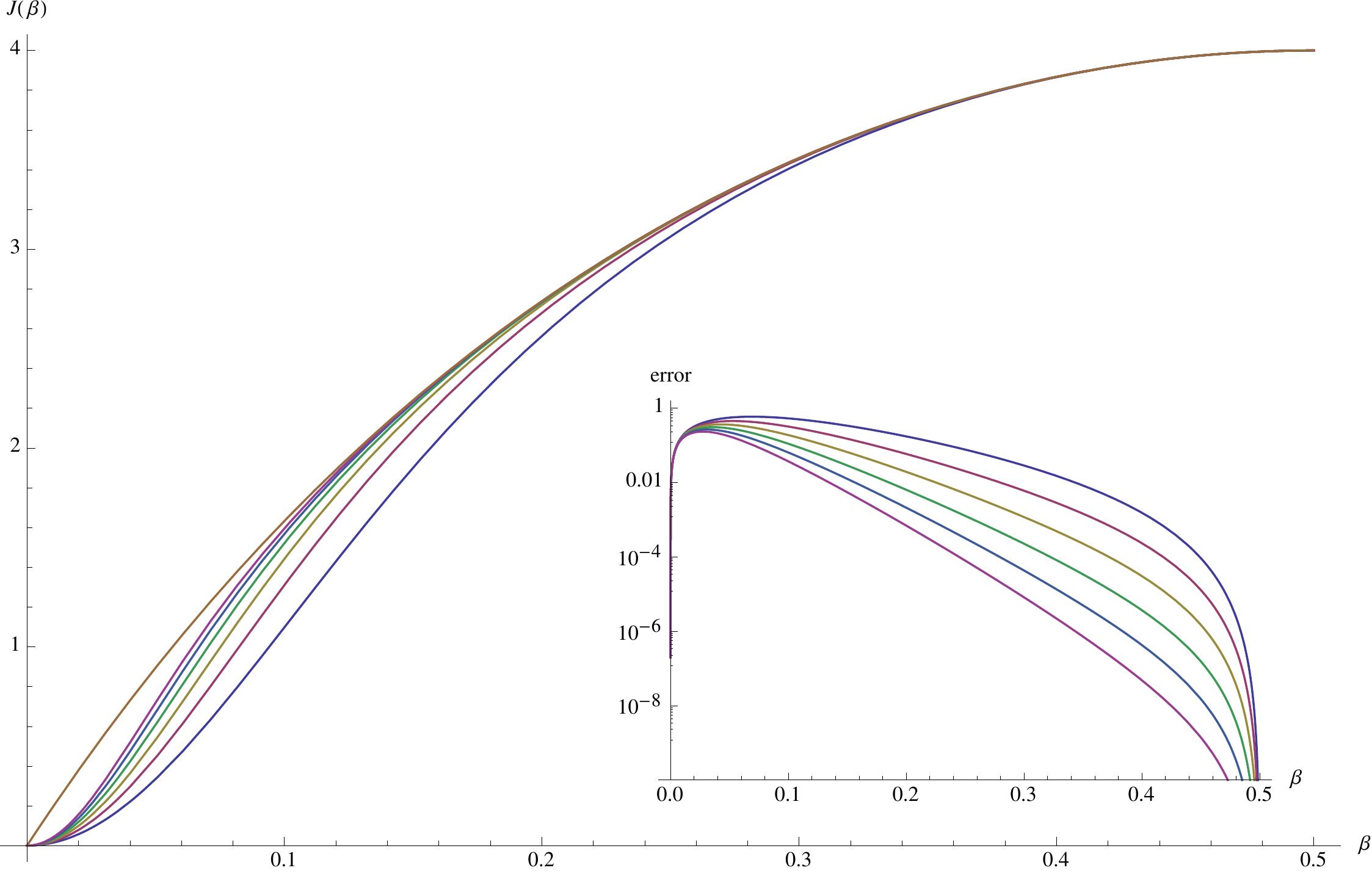}
\hfill
\caption{\label{fig:Jbeta} $J(\beta)$ as given by \eqref{Jbeta2} (top curve) and as calculated by numerical inversion of truncated matrices of dimension $N=2,8,32,128,512,2048$ (lower curves, bottom to top). The functions are invariant under $\beta\to1-\beta$, so only the region $0<\beta<1/2$ is shown. Inset: Difference between \eqref{Jbeta2} and truncated matrix result (same values of $N$, top to bottom).}
\end{figure}

\paragraph{Numerical inversion:} For finite values of $\lambda$, we can approximate $(I+\lambda^2T)^{-1}$ by inverting the truncated matrix, which can be done numerically for rather large matrices. In figure \ref{fig:Jbeta} the resulting approximation to $J(\beta)$ is plotted for matrices up to $2048\times2048$, alongside the function \eqref{Jbeta2}. It can be seen that the approximation is very good for $\beta$ close to 1/2 (i.e.\ $\lambda$ close to 0), even for small matrix sizes. This is to be expected given the fact that the function is analytic there and the fast convergence of the Taylor series coefficients as the matrix size is increased, as discussed above. On the other hand, the convergence is much slower near $\beta=0$. This is presumably related to the fact that we are approximating a non-analytic function by analytic ones.\footnote{Unfortunately, the two meanings of the term ``analytic'' are in conflict in this context: the approximation derived from the truncated matrix, which we are evaluating numerically, is an analytic function, while the analytic expression $\lambda^{-1}\arctan\lambda$ represents a non-analytic function.} Or, to say the same things in terms of the variable $\beta$, we are approximating a function with a finite first derivative by functions with zero first derivative but a large second derivative. Furthermore, as shown above, that second derivative is going to infinity only logarithmically with the matrix size. Nonetheless, it is clear that, for any fixed value of $\beta$, the numerical result is indeed converging to the predicted expression, although the rate of convergence is very slow for small $\beta$.

We believe that, taken together, these three sets of calculations constitute convincing evidence for the correctness of \eqref{Jbeta2}.

\bibliographystyle{JHEP}
\bibliography{refs}

\end{document}